\documentclass[usenatbib]{mn2e}

\usepackage{psfig}

\def\gs{\mathrel{\raise0.35ex\hbox{$\scriptstyle >$}\kern-0.6em\lower0.40ex\hbox{{$\scriptstyle \sim$}}}}
\def\ls{\mathrel{\raise0.35ex\hbox{$\scriptstyle <$}\kern-0.6em\lower0.40ex\hbox{{$\scriptstyle \sim$}}}}

\def\Wm2{\,\hbox{W}\,\hbox{m}^{-2}}
\def\gsim{\mathrel{\raise0.35ex\hbox{$\scriptstyle >$}\kern-0.6em\lower0.40ex\hbox{{$\scriptstyle \sim$}}}}
\def\lsim{\mathrel{\raise0.35ex\hbox{$\scriptstyle <$}\kern-0.6em\lower0.40ex\hbox{{$\scriptstyle \sim$}}}}
\def\ltsima{$\; \buildrel < \over \sim \;$}
\def\simlt{\lower.5ex\hbox{\ltsima}}
\def\gtsima{$\; \buildrel > \over \sim \;$}
\def\simgt{\lower.5ex\hbox{\gtsima}}

\begin{document}

\title[The ISM in a $z\sim 2.3$ Star-Forming Galaxy]{The Properties of
  the Interstellar Medium \\ within a Star-Forming Galaxy at $z$=2.3}

\author[Danielson et al.]
{\parbox[h]{\textwidth}{
A.\,L.\,R.\ Danielson,$^{\,1,* }$
A.\,M.\ Swinbank,$^{\,1}$
Ian Smail,$^{\,1}$
P.\, Cox,$^{\,2}$
A.\,C.\ Edge,$^{\,1}$
A.\ Weiss,$^{\,3}$
A.\,I.\ Harris,$^{\,4}$
A.\,J.\ Baker,$^{\,5}$
C.\ De Breuck,$^{\,6}$
J.\,E.\ Geach,$^{\,1}$
R.\,J.\ Ivison,$^{\,7,8}$
M.\ Krips,$^{\,2}$
A.\ Lundgren,$^{\,9}$
S.\ Longmore,$^{\,10}$
R.\ Neri,$^{\,2}$
\& B.\ Oca\~{n}a Flaquer,$^{\,11}$
}
\vspace*{6pt}\\
$^1$Institute for Computational Cosmology, Department of Physics, Durham University, South Road, Durham DH1 3LE, UK\\
$^2$Institut de Radio Astronomic Millimetrique, 300 rue de la Piscine, Domaine Universitaire, 38406 Saint Martin d'Heres, France\\
$^3$Max-Planck Institut f\"ur Radioastronomie, Auf dem H\"ugel 69, 53121 Bonn, Germany\\
$^4$Department of Astronomy, University of Maryland, College Park, Maryland 20742, USA\\
$^5$Department of Physics and Astronomy, Rutgers, the State University of New Jersey, 136 Frelinghuysen Road, Piscataway, \\ New Jersey 08854-8019, USA\\
$^6$European Southern Observatory, Karl-Schwarzschild Strasse, 85748 Garching bei M\"unchen, Germany\\
$^7$UK Astronomy Technology Centre, Royal Observatory, Blackford Hill, Edinburgh, EH19 3HJ, UK\\
$^8$Institute for Astronomy, University of Edinburgh, Edinburgh, EH19 3HJ, UK\\
$^9$European Southern Observatory, Alonso de Cordova 3107, Cassilla 19001, Santiago 19, Chile\\
$^{10}$Harvard-Smithsonian Center For Astrophysics, 60 Garden Street, Cambridge, Massachusetts 02138, USA\\
$^{11}$Instituto de Radioastronomía Milim\'etrica, Avenida Divina Pastora 7, N\'ucleo Central, 18012 Granada, Spain\\
$^*$Email: a.l.r.danielson@durham.ac.uk\\
}

\maketitle

\begin{abstract}
We present an analysis of the molecular and atomic gas emission in the
rest-frame far-infrared and sub-millimetre, from the lensed $z=2.3$
sub-millimetre galaxy SMM\,J2135$-$0102.  We obtain very high
signal-to-noise detections of 11 transitions from three species and
limits on a further 20 transitions from nine species.  We use the
$^{12}$CO, [C{\sc i}] and HCN line strengths to investigate the gas
mass, kinematic structure and interstellar medium (ISM) chemistry, and
find strong evidence for a two-phase medium within this high-redshift
starburst galaxy, comprising a hot, dense, luminous component and an
underlying extended cool, low-excitation massive component.  Employing
a suite of photo-dissociation region models we show that on average
the molecular gas is exposed to a UV radiation field that is
$\sim1000\times$ more intense than the Milky Way, with star-forming
regions having a characteristic density of
$n\sim10^{4}$\,cm$^{-3}$. Thus, the average ISM density and far-UV
radiation field intensity are similar to those found in local ULIRGs.
These densities and radiation fields are similar to those found in the
central regions of typical starburst galaxies, even though the star
formation rate is far higher in this system.  The $^{12}$CO spectral
line energy distribution and line profiles give strong evidence that
the system comprises multiple kinematic components with different
conditions, including temperature, and line ratios suggestive of high
cosmic ray flux within clouds, likely as a result of high star
formation density. We find tentative evidence of a factor $\sim4$
temperature range within the system. We expect that such internal
structures are common in high-redshift ULIRGs but are missed due to
the poor signal-to-noise of typical observations.  We show that, when
integrated over the galaxy, the gas and star-formation surface
densities appear to follow the Kennicutt-Schmidt relation, although by
comparing our data to high-resolution sub-mm imaging, our data suggest
that this relation breaks down on scales of $<$100\,pc.  By virtue of
the lens amplification, these observations uncover a wealth of
information on the star formation and ISM at $z\sim2.3$ at a level of
detail that has only recently become possible at $z<0.1$, and show the
potential physical properties that will be studied in unlensed
galaxies when the Atacama Large Millimeter Array is in full operation.
\end{abstract}


\section{Introduction}

The properties of the interstellar medium (ISM) play a critical role
in the evolution of galaxies as it includes the raw material from
which stars form.  The main sources of heating of the atomic and
molecular ISM are the stellar radiation field, cosmic rays and
turbulence, and as such the ISM exhibits a considerable range of
properties, spanning a wide range in density $\sim
10^{0-7}$\,cm$^{-3}$ and temperature\,$\sim10$--1000\,K.  The
thermodynamic state of the gas is dictated by the balance of this
heating with cooling from atomic and molecular species (e.g. [C{\sc
    ii}], [C{\sc i}], $^{12}$CO and [O{\sc i}]) and this balance may
dictate the Jeans mass and initial mass function (IMF) and therefore
the efficiency of star formation \citep{Hocuk10}.  Understanding the
balance of heating and cooling within the ISM is thus fundamental to
understanding the detailed physics of the star formation which drives
the formation and evolution of galaxies.

An essential element of our understanding of distant galaxies comes
therefore from observations of emission from interstellar molecular
and atomic gas.  To date, the majority of the detections of $^{12}$CO
and atomic fine structure emission from high-redshift galaxies have
been in powerful QSOs and intrinsically luminous galaxies, typically
with star formation rates $>10^3$\,M$_{\odot}$\,yr$^{-1}$
(e.g.\ \citealt{Frayer98,Frayer99,Greve05,Tacconi06,Tacconi08,HD10,Walter10}).
However, the detailed physical properties of the vigorous starbursts
within high-redshift starburst galaxies are still unknown.  Analogy to
local ultra-luminous infrared galaxies (ULIRGs) would argue for
compact starbursts triggered by mergers, although secular bursts in
massive gas disks are also feasible \citep[e.g.][]{Dave10}.  These
different scenarios imply different properties for the ISM and so may
be tested using observations of the molecular and atomic emission.
Critically, the first full $^{12}$CO spectral line energy
distributions (SLEDs), including high-$J_{\rm upper}$ transitions,
have recently been completed for local ULIRGs \citep{PPP10b, VdW10}
providing an empirical benchmark for comparison to high-redshift
sources, to search for similarities in the thermodynamics of their
ISM.

Much of the atomic and molecular gas emission from galaxies arises
from photo-dissociation regions (PDRs), the surfaces of molecular
clouds where the heating and chemistry is dominated by far-ultraviolet
(far-UV) photons from stellar sources (Hollenbach \& Tielens 1997).
PDRs dominate the infrared and sub-millimetre emission line spectra of
star formation regions and galaxies as a whole.  The majority of their
cooling occurs via atomic fine structure lines of [O{\sc i}], [C{\sc
    ii}] and the rotational $^{12}$CO lines \citep{Kramer04}.

Prior to full science operations of ALMA, the most promising route to
gaining high signal-to-noise, sub-millimetre spectroscopy of
star-forming galaxies at high redshift is to use the natural
amplification from strong gravitational lensing
\citep[e.g.][]{Smail02,Weiss05a,Weiss05b,Weiss07,Maiolino09,HD10,Ivison10L1L2}.
In particular, measuring the chemistry of the ISM through emission
line flux ratios in strongly-lensed galaxies is particularly
advantageous since gravitational lensing is achromatic (in the absence
of differential amplification) and so the apparent line ratios (and
the implied chemistry) are unaffected.  As such, intrinsically faint
emission lines which are sensitive probes of the ISM chemistry (such
as $^{13}$CO or HCN) can be measured, if suitably bright targets can
be identified.

Recently, \citet{Swinbank10Nature} reported the discovery of a
strongly-lensed sub-millimetre galaxy SMM\,J2135$-$0102 (hereafter
SMM\,J2135; R.A./Dec.: 21\,35\,11.6, $-01$\,02\,52; J2000) behind a
massive galaxy cluster at $z=0.325$.  The redshift for the galaxy,
$z=2.3259$, was derived through blind detection of $^{12}$CO(1--0)
using Zpectrometer on Green Bank Telescope (GBT).  Based on a detailed
lensing model, the 870-$\mu$m emission appears to be amplified by a
factor $32.5\pm 4.5\times$, implying an unlensed flux density of $\sim
3$\,mJy (L$_{\rm bol} = 2.3\pm0.4 \times 10^{12}$\,L$_{\odot}$;
\citealt{Ivison10eyelash}), which is close to the confusion limit of
existing sub-millimetre surveys (\citealt{Blain02}).

This uniquely bright source therefore provides an opportunity to
investigate the detailed properties of the ISM within a galaxy which
is representative of the high-redshift starburst population.  In this
paper we analyse follow-up observations of the $^{12}$CO ladder (up to
$J_{\rm upper}=10$), as well as other dense gas tracers and more
complex molecules in the rest-frame frequency range 80--1100\,GHz.  To
put the current molecular/atomic line data obtained for the {\it
  global} line emission of this distant starburst in perspective, we
note that they are rivaled only by those obtained for local ULIRGs,
e.g. Arp\,220, Mrk\,231 and NGC\,6240 \citep{Greve09,VdW10}, which are
some $\sim 250\times$ closer to us.

In \S\ref{sec:obs} we describe our observations of the molecular and
atomic emission from SMM\,J2135. Our analysis is described
in \S\ref{sec:anal} where we first examine the integrated properties of
the system, deriving and comparing gas masses calculated from
$^{12}$CO, [C{\sc i}] and HCN.  We produce a $^{12}$CO SLED for this
galaxy and use this to investigate excitation conditions within its
ISM.  We use PDR models to investigate the characteristic density of
the ISM and the incident far-UV intensity.  We then decompose the
spectra into three distinct kinematic components, and measure the
$^{12}$CO SLED of each in order to search for excitation structure
within the system.  We give our conclusions in \S \ref{conc}.
Throughout the paper we use a $\Lambda$CDM cosmology with
H$_0=72$\,km\,s$^{-1}$\,Mpc$^{-1}$, $\Omega_m=0.27$ and
$\Omega_{\Lambda}=1-\Omega_m$ \citep{Spergel04}.  Unless otherwise
stated, a lensing amplification correction of a factor of 32.5 has
already been applied to all quoted luminosities.
\section{Observations}
\label{sec:obs}

%
% Table 1
%
\begin{table}
\begin{center}
{\small
\centerline{\sc Table 1.}
\centerline{\sc Log of IRAM 30\,m and GBT Observations}
\smallskip
\begin{tabular}{cclc}
\hline
\noalign{\smallskip}
 Band   & $\nu_{\rm obs}$ & ~\hspace*{0.5in} Emission lines   & t$_{\rm int}$ \\
        &  (GHz) &                                         & (ks) \\
\hline
Ka        & 25.6--36.1      & $^{12}$CO(1--0), CS(2--1), CN(1--0)  \\
 & & HNC(1--0), $^{13}$CO(1--0) \\
 & & HCO$^{+}$(1--0) & 18.0 \\
\noalign{\smallskip}
E0       & 79.00--83.00    & HCN(3--2), HCO$^+$(3--2), & \\
& & HNC(3--2)                & 28.8 \\
        & 96.75--100.75   & $^{13}$CO(3--2), H$_2$O\,325.141                 & 18.0 \\
        & 101.97--105.97  & $^{12}$CO(3--2), CN(3--2), CS(7--6)  & 18.0 \\
\noalign{\smallskip}
E1        & 131.00--135.00   & $^{13}$CO(4--3), HCO$+$(5--4),& \\
& & HCN(5--4), CS(9--8) & 14.4 \\
        & 135.92--139.92  & $^{12}$CO(4--3), CN(4--3),  HNC(5--4) & 10.8 \\
        & 146.00--150.00  & [C{\sc i}](1--0), CS(10--9),  O$_2$\,487.3        & ~7.2  \\
        & 171.26--174  & $^{12}$CO(5--4)                               & 10.8 \\
\noalign{\smallskip}
E2        & 205.90--209.90  & $^{12}$CO(6--5)                               & ~8.4  \\
        & 224.11--228.11  & H$_2$O\,731.681                              & ~3.6 \\
        & 240.53--244.53  & $^{12}$CO(7--6), [C{\sc i}](2--1)                & 12.6 \\
\noalign{\smallskip}
E3       & 275.16--279.16  & $^{12}$CO(8--7)                               & ~9.6 \\
        & 309.77--313.77  & $^{12}$CO(9--8)                               & ~7.2 \\
\hline
\label{tab:EMIRsetup}
\end{tabular}
}
\end{center}
\footnotesize{

The log of the IRAM 30m and GBT Zpectrometer observations giving the
frequency ranges of Zpectrometer Ka band and the IRAM/EMIR receiver
backends (E0--E3), the emission lines observed and integration time
for each set-up.\\
Note: These set-ups have beam sizes of
Ka:$\sim15$\arcsec--23\arcsec, E0:$\sim30$\arcsec, E1:$\sim17$\arcsec,
E2:$\sim11$\arcsec and E3:$\sim8.5$\arcsec respectively.}

\end{table}

%
% Table 2
%

\begin{table}
\begin{center}
\small 
\centerline{\sc Table 2.} 
\centerline{\sc Emission Line Properties} 
\smallskip
\begin{tabular}{lccc}
\hline
\noalign{\smallskip}
Species          & $\nu_{\rm rest}$ & Flux$^{a,b}$          & L$'$\\
                  & (GHz) & (Jy\,km\,s$^{-1}$)  & (10$^8$\,K\,km s$^{-1}$\,pc$^{2}$)  \\
\hline
$^{12}$CO(1--0)     & ~115.2712  & $2.16\pm 0.11$    & $173\pm9$        \\
$^{12}$CO(3--2)     & ~345.7959  & $13.20\pm 0.10$ & $117.6\pm0.9$        \\
$^{12}$CO(4--3)     & ~461.0408  & $17.3\pm 1.2$ & $87\pm6$     \\
$^{12}$CO(5--4)     & ~576.2679  & $18.7\pm 0.8$ & $60\pm 3$     \\
$^{12}$CO(6--5)     & ~691.4731  & $21.5\pm 1.1$ & $48 \pm2$     \\
$^{12}$CO(7--6)     & ~806.6518  & $12.6\pm 0.6$ & $21\pm 1$     \\
$^{12}$CO(8--7)     & ~921.7997  & $8.8\pm 0.5$  & $11\pm1$     \\
$^{12}$CO(9--8)     & 1036.9124 & $3.6\pm 1.7$  & $4\pm 2$     \\
$^{12}$CO(10--9)    & 1151.9855 & $<1.1$        & $<0.9$            \\
\noalign{\smallskip}
$^{13}$CO(1--0)     & ~110.2013  & $<0.3$        & $<29$          \\
$^{13}$CO(3--2)     & ~330.5880  & $<1.8$        & $<18$           \\
$^{13}$CO(4--3)     & ~440.7651  & $<0.3$        & $<1.4$           \\
\noalign{\smallskip}
$[$C{\sc i}$]$(1--0)     & ~492.1607  & $16.0\pm 0.5$ & $71\pm 2$     \\
$[$C{\sc i}$]$(2--1)     & ~809.3435  & $16.2\pm 0.6$ & $26\pm 1$     \\
\noalign{\smallskip}
HCN(1--0)          & ~~88.6300   & $<0.3$        & $<45$          \\
HCN(3--2)          & ~265.8900  & $1.2\pm0.4$   & $18\pm6$     \\
HCN(5--4)          & ~443.1200  & $<0.3$        & $<1.4$         \\
\noalign{\smallskip}
HNC(1--0)          & ~~90.6600   & $<0.3$        & $<43$          \\
HNC(3--2)          & ~271.9800  & $<1.2$         & $< 17$   \\
\noalign{\smallskip}
HCO$+$(1--0)         & ~~89.190   & $<0.3$        & $<44$          \\
HCO$+$(3--2)         & ~267.560  & $<1.2$        & $<17$           \\
HCO$+$(5--4)         & ~445.90~  & $<0.3$        & $<1.3$           \\
\noalign{\smallskip}

CN(1--0)           & ~113.320  & $<0.3$       & $<27$      \\
CN(3--2)           & ~339.450  & $<1.6$       & $<14$            \\
\noalign{\smallskip}
CS(2--1)           & ~~97.981   & $<0.3$       & $<37$           \\
CS(7--6)           & ~342.883  & $<1.6$       & $<14$            \\
CS(9--8)           & ~428.875  & $<0.3$       & $<14$            \\
CS(10--9)          & ~489.751  & $<1.6$        & $<7$            \\
\noalign{\smallskip}
O$_2$              & ~487.246  & $<1.6$        & $<7$            \\
\noalign{\smallskip}
H$_2$O            & ~325.141  & $<1.8$        & $<18$           \\
H$_2$O            & ~731.681  & $<4.8$        & $<9.5$            \\
\noalign{\smallskip}
[C{\sc ii}]$^{c}$             & 1910~~~~  & $850\pm180$      &  $2500\pm500$            \\
\noalign{\smallskip}
[OI]$^{d}$              & ~2065.40  &  $620\pm200$   &   $155\pm50$ \\ 
\hline
\label{tab:photom}
\end{tabular}
\end{center}
\footnotesize{

$^a$We quote 3-$\sigma$ limits for all lines which are not formally
  detected.\\ $^b$Uncertainties on fluxes denote measurement errors
  and do not include the calibration uncertainties, which we estimate
  as $\sim5$\% for 30--200\,GHz, $\sim10$\% for 200--300\,GHz and
  $\sim15$\% for $>300$\,GHz.\\ $^c$The [C{\sc ii}] flux is taken from
  \citet{Ivison10eyelash}. \\ $^{d}$\cite{Ivison10eyelash} report a
  $\sim3\sigma$ from [OI]146$\mu$m and we have used the flux of this
  feature in our analysis, although we note that removing this line
  from our analysis changes our results negligibly. We also note that
  the since the [C{\sc ii}] and [O{\sc i}] 146$\mu$m lines are of
  comparable strength, this would imply that the [O{\sc i}] 63$\mu$m
  line would be much brighter than the [C{\sc ii}] line thus making
  this an interesting source for Herschel/PACS follow-up. }

\end{table}

An essential requirement for the study of the molecular and atomic
emission lines from SMM\,J2135 is a precise redshift for the gas
reservoir in this system.  This was obtained soon after the discovery
of the source using Zpectrometer on GBT (\citealt{Swinbank10Nature},
see \S\ref{zpec}).  With this accurate systemic redshift, it was then
possible to precisely tune to the expected frequencies of molecular
and atomic emission lines.

\subsection{GBT Zpectrometer Observations}
\label{zpec}

Details of the observations with Zpectrometer are given in
\citet{Swinbank10Nature} and Table 1.  Briefly, Zpectrometer is a
wide-band spectrometer optimised for $^{12}$CO(1--0) emission line
searches between $z =2.2$--3.5 (with $\sim 150$\,km\,s$^{-1}$
resolution) using the GBT's Ka-band receiver (see \citealt{Harris06}
and \citealt{Harris10} for more detail on the instrument and observing
mode).  Observations of SMM\,J2135 were conducted in two equal shifts
on 2009 May 19 and May 27 for a total integration time of 5\,hours.
Data reduction was carried out with the standard Zpectrometer GBT
reduction scripts. The final spectrum reaches an rms noise level of
$\sigma= 0.50$\,mJy\,beam$^{-1}$.  From the spectrum, shown in
Fig.~\ref{fig:all_spec}, we determine the heliocentric redshift as $z
=2.32591$ from the $^{12}$CO(1--0) emission at 34.648\,GHz.  The
observed velocity-integrated flux in $^{12}$CO(1--0) is given in Table
2.

\subsection{IRAM PdBI \& 30m Observations}

We used the six-element IRAM Plateau de Bure Interferometer (PdBI) to
observe the redshifted $^{12}$CO(3--2) and $^{12}$CO(4--3) emission
lines and the continuum at 104\,GHz and 139\,GHz
respectively. Observations were made in D configuration in Director’s
Discretionary Time (DDT) on 2009 May 29 and 31 with good atmospheric
phase stability and precipitable water vapour (seeing = 0.6--$1.6''$,
pwv\,=\,5--15mm).  The receivers were tuned to the systemic redshift
determined from the GBT $^{12}$CO(1--0) spectrum.  We observed
SMM\,J2135 for total on-source exposure times of 4\,hrs and 2\,hrs for
$^{12}$CO(3--2) and $^{12}$CO(4--3) respectively.  The correlator was
adjusted to a frequency resolution of 2.5\,MHz, yielding 980-MHz
coverage.  The overall flux scale for each observing epoch was set
from observations of MWC\,349, with additional observations of
2134+004 for phase and amplitude calibrations.  The data were
calibrated, mapped and analyzed using the {\sc gildas} software
package.  Inspection of the velocity datacubes shows very good
detections of both $^{12}$CO(3-–2) and $^{12}$CO(4--3) emission lines
(S/N\,$\sim300$ in each) at the position of SMM\,J2135 (see
Fig.~\ref{fig:all_spec}) and we give line fluxes in Table~2.

To constrain the high-$J_{\rm upper}$ $^{12}$CO emission (and search
for other emission lines such as [C{\sc i}]) we used the Eight MIxer
Receiver (EMIR) multi-band hetrodyne receiver at the IRAM 30-m
telescope.  Observations were made on 2009 June 29--30, and 2010
February 5--8 and April 3--6, in good to excellent conditions,
typically with $\sim 2$--6\,mm pwv and $\lsim 1$\,mm pwv for some of
the April observations.  Data were recorded using the E0--E3 receivers
with 4\,GHz of instantaneous, dual-polarisation band-width covering
frequency ranges from $\sim 70$--310\,GHz (see Table~1 for the details
of the setups).  For each observation, we used eight 1-GHz bandwidth
units of the Wide-band Line Multiple Autocorrelator (WILMA) to cover
4\,GHz in both polarizations.  WILMA provides a spectral resolution of
2\,MHz which corresponds to 5--7\,km\,s$^{-1}$ in the 3-mm band. The
observations were carried out in wobbler-switching mode, with a
switching frequency of 1\,Hz and an azimuthal wobbler throw of
$90''$. Pointing was checked frequently on either the nearby quasar
J2134+004 or Venus and was found to be stable to within $3''$.
Calibration was carried out every 12\,mins using the standard
hot/cold-load absorber, and the flux calibration was carried out using
the point source conversion between temperature and flux as measured
from celestial objects.  We note that, even at the highest frequency
(i.e.\ 345\,GHz) the beam is $8.5''$ and so the lensed galaxy is
likely to be smaller than the beam at all wavelengths
\citep{Swinbank10Nature}.  The data were first processed with the {\sc
  class} software \citep{Pety05}, and then using custom {\sc idl}
routines.  We omitted scans with distorted baselines and subtracted
only linear baselines from individual spectra.  For the 2010 April
observations of $^{12}$CO(5--4), an upward correction was made to the
flux of 10\% since the line extends into the wings of an atmospheric
water band (although for these observations the pwv $\lsim 1$\,mm).
In total, we observed twelve different set-ups, each for $\sim
2$--5\,hrs (Table~1); the spectra of the detections are shown in
Fig.~\ref{fig:all_spec}.  These observations provide detections or
limits on 31 transitions listed in Table 2.

\subsection{APEX SHFI Observations}

Observations of the redshifted $^{12}$CO(7--6) emission line were also
carried out in DDT using the Swedish Hetrodyne Facility Instrument
(SHFI) between 2009 July 10 and July 20 (ESO programme ID 283.A-5014).
SHFI consists of four wide-band heterodyne receiver channels for
230--1300\,GHz (\citealt{Vassilev08}).  We used APEX-1 tuned at
242.4\,GHz to search for the $^{12}$CO(7--6) emission from the galaxy,
and obtained a total integration time of $\sim 5.5$\,hrs in excellent
conditions (0.4--0.6\,mm pwv).  The data reduction was carried out
using the {\sc class} software, omitting scans with poor baselines.
To create the final spectrum we binned the data onto a velocity scale
of 50\,km\,s$^{-1}$ and averaged it together with 4\,hours of
observations from IRAM/EMIR, which has similar signal-to-noise; we
show this spectrum in Fig.~\ref{fig:all_spec}.

\subsection{SMA Observations}

\begin{figure*}
\centerline{
\psfig{figure=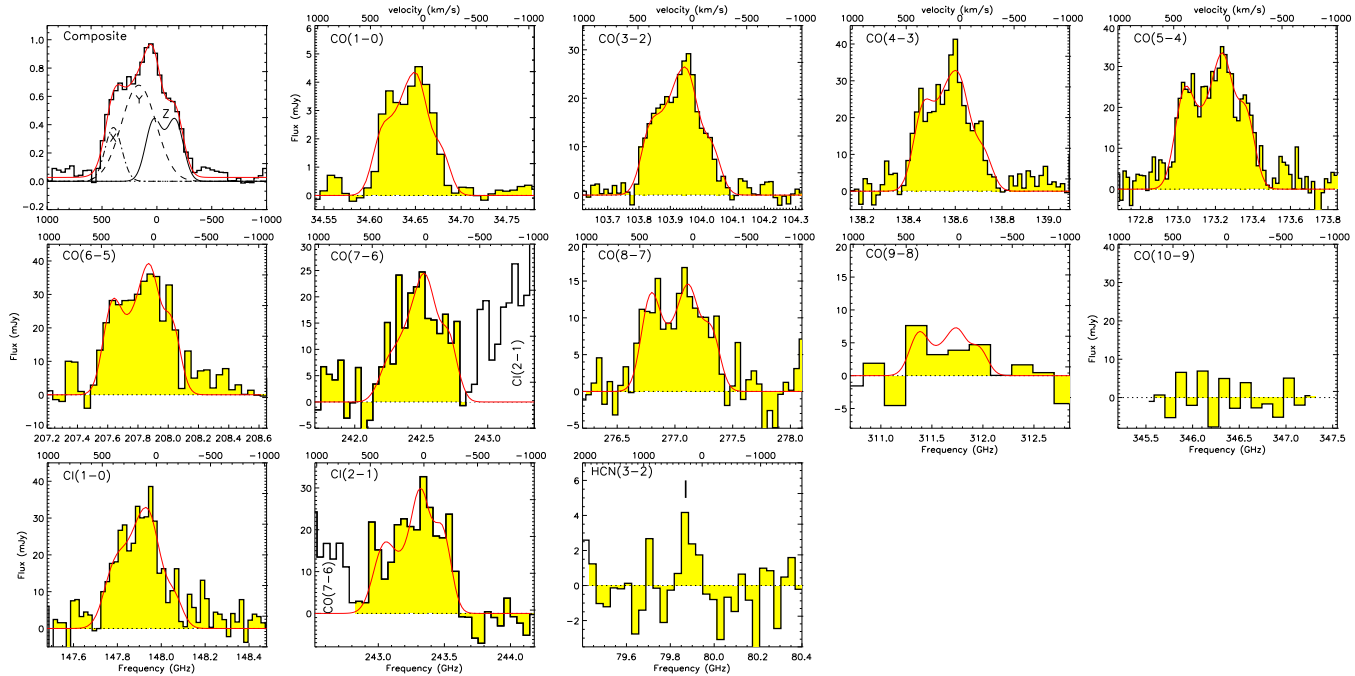,width=7in,angle=90}}
\caption{Molecular and atomic emission from the lensed sub-millimetre
  galaxy SMM\,J2135. The top two rows show spectra of the $^{12}$CO
  emission arising from $J_{\rm upper}=1$ up to $J_{\rm upper}=10$.
  These are followed by the spectra of the [C{\sc i}] fine-structure
  lines and our detection of HCN(3--2).  In all cases, the emission
  profiles show multiple velocity components whose intensity appears
  to vary between transitions.  To better constrain the kinematic
  structure of the lines we derive an average $^{12}$CO spectrum
  (which does not include $^{12}$CO(9--8) or $^{12}$CO(10--9)) and we
  overplot on this the resulting three-component parametric model as
  described in \S\ref{sec:dec_cosled}. We fit this three-component
  kinematic model to the various lines, allowing the intensities of
  the components to vary between lines, and overlay this on all the
  $^{12}$CO emission lines making up the average, clearly highlighting
  the evidence of different excitation in the different velocity
  components. The spectra have been binned into channels of width
  30--100 km s$^{-1}$.  The spectra for $^{12}$CO(3--2),
  $^{12}$CO(4--3) and $^{12}$CO(6--5) are the combined data from our
  PdBI and EMIR observations, the $^{12}$CO(7--6) spectrum is a
  combination of EMIR and SHFI data. The $^{12}$CO(1--0) line was measured
  by GBT and the $^{12}$CO(10--9) line by SMA, all other observations were
  taken with IRAM 30m and PdBI. The axes are flux (mJy) and frequency
  (GHz) (with the velocity shown on the upper horizontal axes) in all
  cases except for the average spectrum which has been plotted as
  km\,s$^{-1}$ and flux normalised to the peak value. Note that
  $^{12}$CO(7--6) and [C{\sc i}](2--1) lines abut each other in the
  spectrum.  }
\label{fig:all_spec}
\end{figure*}

We mapped the 870-$\mu$m continuum emission from SMM\,J2135 using the
Submillimeter Array (SMA) in a number of different array
configurations (see \citealt{Swinbank10Nature}). During four of these
tracks (in the sub-compact, compact, extended and very-extended
configurations) we tuned the upper side band of the receiver to
346.36699\,GHz to search for $^{12}$CO(10--9).  Details of the data
reduction are given in \citet{Swinbank10Nature} and we show the
spectrum in Fig.~\ref{fig:all_spec}.  The resulting combined spectrum
reaches an rms of 2\,mJy per 100\,km\,s$^{-1}$ channel and although
$^{12}$CO(10--9) is not detected, we place a 3-$\sigma$ limit on the
flux of $\leq 0.2$\,Jy\,km\,s$^{-1}$.

%
% Figure 1
%

\section{Analysis and Discussion}
\label{sec:anal}

Our observations have detected 11 individual transitions arising from
three molecular or atomic species in SMM\,J2135 and place upper limits
on the line fluxes of a further 20 transitions arising from seven other
species.  We list these in Table~2 with the fluxes quoted with their
respective measurement uncertainties and we show the spectra for all
detections in Fig.~\ref{fig:all_spec}.  We estimate additional
calibration uncertainties of 5\%, 10\% and 15\% for those lines at
30--200\,GHz, 200--300\,GHz and $>300$\,GHz respectively.

With the very high signal-to-noise of our data (particularly the
$^{12}$CO lines), we clearly identify multiple velocity components
within the spectra (at approximately $-180$, 0 and
$+200$\,km\,s$^{-1}$ relative to the $^{12}$CO(1--0) flux-weighted
mean velocity), implying multiple physical components in the source.
We believe that the detection of this structure in the $^{12}$CO
spectra is not due to this galaxy's being unusual, but instead
reflects the high signal-to-noise of our observations, which is
uncommon for observations of other (even local) galaxies. We discuss
these multiple components in \S\ref{sec:dec_cosled}, modelling the
system as three components which we label {\it X}, {\it Y} and {\it
  Z}.  However, so that fair comparisons can be made to other galaxies
(at low and high redshift), we begin (in \S \ref{sec:Mgass},
\ref{sec:cosled} and \ref{sec:PDRs}) by discussing the integrated
properties of the system to see what can be learnt about the bulk
properties of its ISM.

As Fig.~\ref{fig:all_spec} shows, the $^{12}$CO and [C{\sc i}]
emission lines are broad, with a typical FWHM of $\sim
500$\,km\,s$^{-1}$.  Since there is significant kinematic structure
within the $^{12}$CO and [C{\sc i}] emission, we first define the
width of the emission lines by constructing a composite $^{12}$CO
spectrum by combining the spectra from $^{12}$CO(1--0) to
$^{12}$CO(8--7) (normalised by peak flux).  This composite has a full
width at zero intensity of 900\,km\,s$^{-1}$, with a range from $-350$
to 550\,km\,s$^{-1}$ (Fig.~\ref{fig:all_spec}), although we note that
there may be faint emission extending to $\pm1000$km s$^{-1}$.  We use
the velocity range, $-350$ to 550\,km\,s$^{-1}$, to measure the fluxes
of the $^{12}$CO lines, [C{\sc i}]($^3P_1\rightarrow\,^3P_0$) and
[C{\sc i}]($^3P_2\rightarrow\,^3P_1$), while for HCN(3--2) we use a
range of $-100$ to 550\,km\,s$^{-1}$.  To determine the errors on the
fluxes we measure the variance in the spectra away from the emission
line, although in most cases the dominant error is from the
calibration; these errors are used in Fig.~\ref{fig:CO_SLED_tot} (see
also Table~2).

\subsection{Integrated Properties}
\subsubsection{Gas Mass and Dust SED}
\label{sec:Mgass}

We begin by estimating the total molecular gas mass of the galaxy from
$^{12}$CO. Low-$J_{\rm upper}$ transitions of $^{12}$CO are commonly
used to obtain estimates of the total gas mass of a galaxy using a
conversion factor, $\alpha=$\,M$_{\rm H_{2}}$\,/\,L$'_{\rm
  ^{12}CO(1-0)}$\,M$_{\odot}$\,(K\,km\,s$^{-1}$\,pc$^2$)$^{-1}$ which
converts the $^{12}$CO line luminosity to total gas mass, (where
M$_{\rm H_2}$ is defined to include the mass of Helium such that
M$_{\rm H_2}=$\,M$_{\rm gas}$; see \citealt{SolomonVandenBout05} for a
review).

There is considerable uncertainty about the correct value of $\alpha$
to adopt in high-redshift galaxies
(e.g.\ \citealt{Baker04,Coppin07,Tacconi08} and Ivison et al.\ 2010c)
and similar uncertainty about the conversion factor to transform
high-J$_{\rm upper}$ CO transitions to the equivalent CO(1--0) fluxes
(if required). However, we can estimate a minimum mass (and so a lower
limit on $\alpha$) assuming the $^{12}$CO(1--0) emission is optically
thin (which is unlikely to be the case) and has solar abundance,
following \citet{Ivison10L1L2}:
\begin{eqnarray*}
 \frac{M(\rm H_{2})}{L'_{\rm CO(1-0)}} & \sim &  0.08\left(\frac{g_1}{Z}e^{-T_o/T_k}\left(\frac{J(T_k)-J(T_{bg})}{J(T_k)}\right)\right)^{-1} \\
& & \times \left(\frac{[\rm ^{12}CO/H_2]}{10^{-4}}\right)^{-1}\frac{M_{\odot}}{\rm K km s^{-1}pc^2}
\end{eqnarray*}
\smallskip
where $T_o=E_{1}/k_{B}\sim5.5$\,K, $J(T)=T_{o}(e^{T_o/T}-1)^{-1}$,
$T_{\rm bg}=(1+z)T_{\rm CMB}\sim9.1$\,K ($T_{\rm CMB}=2.73$\,K at
$z=0$), $z=2.3259$, $g_1=3$ (the degeneracy of level $n=1$),
$Z\sim2(T_k/T_0)$ and [$^{12}$CO/H$_{2}$]$\sim10^{-4}$ for a
solar metallicity environment (\citealt{Bryant96}).  For typical
star-forming gas, T$_{\rm k}\sim40$--60\,K, so to derive a
minimum mass we use T$_{\rm k}=40$\,K. With L$'_{\rm ^{12}CO(1-0)}=173
\pm 9\times 10^{8}$\,K\,km\,s$^{-1}$\,pc$^2$, this gives us a minimum
gas mass of M$_{\rm gas}\gsim1\times10^{10}$\,M$_{\odot}$ and a lower
limit on $\alpha\gsim 0.54$.

We can then compare this to the dynamical mass adopting M$_{\rm
  dyn}$=5\,$R \sigma^2/G$ \citep{SolomonVandenBout05}.  This assumes
the dynamics of the CO emission trace the virialised potential well of
the system and that the CO is distributed in a sphere of constant density
with $r\sim1$\,kpc (see \S \ref{sec:structure} and
\citealt{Swinbank10Nature}). Using the width of the $^{12}$CO(1--0)
line, $\sigma\sim 200$\,km\,s$^{-1}$, we estimate a dynamical mass of
M$_{\rm dyn}\sim5.3\times10^{10}$\,M$_{\odot}$.  If this mass is
dominated by gas, then the upper limit on $\alpha$ is $\alpha=M_{\rm
  gas}/L'_{\rm ^{12}CO(1-0)}<3$. \\

Using this dynamical mass and the minimum gas mass, we place a lower
limit on the gas fraction of M$_{\rm gas}$\,/M$_{\rm dyn} \gsim 0.19$.
This ratio is similar to the ratio for typical starburst nuclei and
the ratio of 30\% found in M\,82 \citep{Dev94}. Compared to other
high-redshift sub-millimetre galaxies (SMGs), using the $^{12}$CO(3--2)
observations from \cite{Greve05} and adopting $\alpha=0.8$ and
R$_{3,1}=0.58\pm 0.05$ from Ivison et al.\ (2010c), we find a median gas
mass ratio of 50\% for luminous SMGs, whilst
\citet{Tacconi08} derive a gas mass fraction of $0.3$ for more typical
star-forming galaxies at similar redshifts. We stress, however, that
compared to the gas mass of $6\times 10^{8}$\,M$_{\odot}$ within the
central 1.2\,kpc of M\,82 \citep{Young84}, our gas mass is nearly two
orders of magnitude higher, underlining the more extreme conditions in
the central regions of SMM\,J2135.

In order to provide a simple comparison with previous studies, we
estimate the gas mass of SMM\,J2135 using a conversion factor of
$\alpha\sim0.8$ which applies to the smoothly distributed,
high-pressure, largely molecular ISM measured in local ULIRGs
(\citealt{Downes98,SolomonVandenBout05}, although recent local studies
on high-J$_{\rm upper}$ CO lines imply multi-phase ISM in ULIRGs as
opposed to smoothly distributed) and is also the canonical value used
for high-redshift LIRGs and ULIRGs \citep{Tacconi08,Stark08}.  This
value of $\alpha$ is comfortably between the upper and lower limits
calculated above. With this assumption the $^{12}$CO(1--0) line
luminosity yields a total gas mass of M$_{\rm H_2}=(1.4\pm0.1)
\times10^{10}$\,M$_{\odot}$ (see also \citealt{Swinbank10Nature}).

Taking our estimate of the total gas mass and the expected size of
$\sim 1$\,kpc for the system \citep{Swinbank10Nature}, we derive an
average column density of $\sim10^{24}$\,cm$^{-2}$, which is
comparable to the molecular hydrogen density in Arp\,220, averaged
over a similar region \citep{Gerin98}, but much higher than the
density in typical starburst galaxies such as M\,82.  However, this
estimate gives an average over the system which we know to be
structured and so indicates that the column density is very high in
parts of this source. The associated extinction is expected to exceed
A$_{\rm V}\sim10^3$, suggesting significant absorption even in the
far-infrared and moreover that the emission from some of the more
common species we see is optically thick.  With this extinction,
coupled with our high value for G$_{0}$ (derived in \S
\ref{sec:PDRmodels}), high infrared luminosity and potentially high
cosmic ray flux (see below), it is probable that momentum-driven
outflows will result. These outflows could be either photon-driven
\citep{Thompson05} or cosmic ray-driven \citep{Socrates08} and will
impact both the dynamics of the gas in the system and the steady-state
assumption in our PDR modelling (\S \ref{sec:PDRmodels}).

We can now compare the star-formation efficiency in SMM\,J2135 to that
in the wider SMG and ULIRG populations.  First we note that the
far-infrared luminosity of SMM\,J2135 is L$_{\rm
  FIR}=(2.3\pm0.1)\times10^{12}$\,L$_{\odot}$ \citep{Ivison10eyelash}
which indicates a star formation rate (SFR) of
$\sim400\pm20$\,M$_{\odot}$yr$^{-1}$ \citep{KS98} assuming a Salpeter
IMF.  We combine the star-formation rate and gas mass to estimate the
star formation efficiency (SFE) following \citet{Greve05},
SFE\,$=L_{\rm FIR}/M_{\rm H_2}$ resulting in
SFE\,$\sim165\pm7$\,L$_{\odot}$\,M$_{\odot}^{-1}$.  This is
comfortably within the limit derived by \citet{Scoville04} of [L$_{\rm
    FIR}/M_{\rm H_2}$]$_{\rm max}
=500$\,L$_{\odot}$\,M$_{\odot}^{-1}$, assuming Eddington-limited
accretion of gas onto OB star clusters.  $^{12}$CO(1--0) measurements
in local ULIRGs derive a SFE of
$180\pm160$\,L$_{\odot}$\,M$_{\odot}^{-1}$, which is comparable to
that in our source.  The SFE of our source is also comparable to the
median SFE of $210\pm80$\,L$_{\odot}$\,M$_{\odot}^{-1}$ for the sample
of luminous SMGs of \cite{Greve05}, after correcting the latter using
$R_{3,1}=0.58$ as appropriate for SMGs (\citealt{Harris10}; Ivison et
al.\ 2010c).

We begin modelling this system by noting that the 870-$\mu$m SMA
observations and comparison to the LABOCA flux reveal four bright,
compact clumps embedded in a much more extended system with the clumps
emitting $\sim80$\% of the total luminosity, and the rest of the
emission emerging from a more extended component
\citep{Swinbank10Nature}.  Motivated by this structure we fit a
two-component model to the dust SED, fixing the characteristic
temperature of the cool (presumably more extended) component at 30\,K
and allowing the size and the characteristic temperature of the clumps
to vary.  The best fit yields a characteristic size for the clumps of
$r\sim 200$\,pc at a temperature of $T_{\rm d,warm}=57\pm3$\,K, while
the extended component has a size of $r\sim1000$\,pc at T$_{\rm
  d,cool}=30$\,K. The dust mass of both the extended component and the
clumps is then $(1.5\pm0.2)\times 10^{8}$\,M$_{\odot}$, which is
similar to the mass determined for the cool dust component in
\cite{Ivison10eyelash}.  The inferred clump size is somewhat larger
than the size of the dust emission regions seen in the highest
resolution 870-$\mu$m SMA maps, where the clumps appear to be
100--200\,pc in diameter \citep{Swinbank10Nature}.  However, we note
that forcing the clump sizes down to that size requires an increase in
the characteristic temperature up to $\sim80$\,K, and to keep the
emission consistent with the integrated SED the clumps must then be
optically thick at $\sim 200\mu$m in the rest-frame.  Clearly this
two-component model is not a unique description of the dust emission
from SMM\,J2135, but it does suggest that the system could be modelled
by a combination of warm, compact and cooler, more extended dust
component.

\subsubsection{$^{12}$CO SLED}
\label{sec:cosled}
%
% Figure 2
%
\begin{figure*}
\centerline{
  \psfig{figure=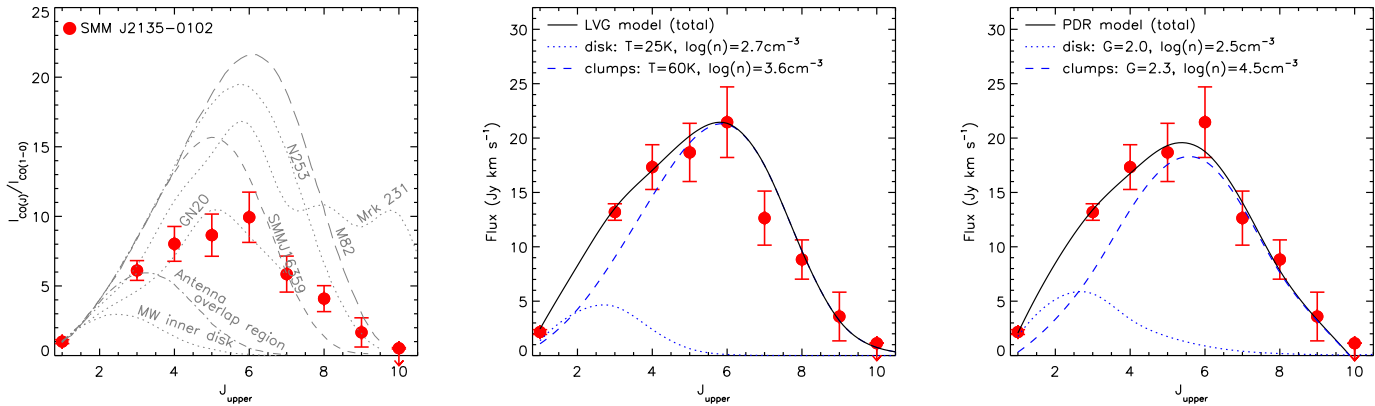,width=7.5in,angle=90}}
\caption{The integrated $^{12}$CO SLED for SMM\,J2135 showing that the
  SLED peaks around $J_{\rm upper}=6$ (similar to M\,82, but with
  proportionally stronger $^{12}$CO(1--0)). The central panel shows
  the results of LVG modelling applied to the integrated SLED, which
  requires two temperature phases to yield an adequate fit.  In this
  model they have characteristic temperatures of T$_{\rm kin}=25$\,K
  and 60\,K and densities of $n=10^{2.7}$ and $10^{3.6}$\,cm$^{-3}$
  respectively (we associate these two phases with a cool, extended
  disk and hotter, more compact, clumps).  The right-hand panel shows
  a similar comparison but now using a PDR model (\citealt{Meijerink07};
  see \S \ref{sec:PDRmodels}) which again requires a combination of
  both low- and high-density phases to adequately fit the integrated
  SLED.  }
\label{fig:CO_SLED_tot}
\end{figure*}

To investigate the excitation within the ISM, we next construct the
$^{12}$CO SLED by calculating the $^{12}$CO flux of each transition
over the velocity range defined by the composite spectrum; we show
the resulting SLED in Fig.~\ref{fig:CO_SLED_tot}.  In Table~4 we also
give the velocity-averaged brightness temperature ratios between
$^{12}$CO($J_{\rm upper}$) and $^{12}$CO(1--0), which we define as
$R_{J,1}=L'(J+1 \rightarrow J)/L'(1 \rightarrow 0)$
(\citealt{Greve03,Harris10}).  The $^{12}$CO SLED shows a continuous
rise out to $J_{\rm upper}=6$, followed by a sharp decline as far as
$J_{\rm upper}=10$. The brightness ratios of all of the $J_{\rm
  upper}>2$ line luminosities compared to the $^{12}$CO(1--0)
are $<1$, which interpreted in the framework of a single phase ISM
would indicate that the gas is sub-thermalised (the thermalised
prediction is $\sim1$; \citealt{Dev94}). Indeed the $L'_{\rm
 ^{12}CO(3-2)}/L'_{\rm ^{12}CO(1-0)}$ line luminosity ratio for
SMM\,J2135, $0.68\pm 0.03$, is similar to that seen in local starburst
galaxies, $0.64\pm0.06$ \citep{Dev94}. However, as \citet{Harris10}
discuss and as we show later in \S\ref{sec:PDRs}, these ratios are
more likely to indicate a multi-phase ISM as found in star-forming
regions locally.

The peak of the SLED at $J_{\rm upper}=6$ is similar to that seen in
nearby starburst galaxies such as NGC\,253 and M\,82
(\citealt[][]{Bradford03, Weiss05c, Panuzzo10}) and AGN,
e.g.\ Mrk\,231 \citep{PPP07,VdW10}.  However, the overall shape of the
SLED suggests that SMM\,J2135 has proportionally stronger
$^{12}$CO(1--0) emission (compared to the higher $J_{\rm upper}$
transitions) than any of these galaxies, indicating the presence of an
additional low-excitation gas phase.  Nevertheless, the bulk of the
cooling, 60\%, arises through the $^{12}$CO J$_{\rm upper}=5$--7 line
emission, with a further 20\% at $^{12}$CO J$_{\rm upper}\geq 8$.
\cite{VdW10} have recently constrained the $^{12}$CO in Mrk\,231 up to
$J_{\rm upper}=13$ with {\it Herschel}, building upon the earlier
study of \cite{PPP07}, and show that, unlike nearby starbursts
(e.g.\ NGC\,253 or M\,82), the $^{12}$CO line luminosity is roughly
flat across J$_{\rm upper}=5$--13. They calculate that only 4\% of the
total $^{12}$CO line luminosity arises from the three lowest
transitions, compared to $\sim20$\% in the three lowest transitions of
our $^{12}$CO SLED (and 43\% in the Milky Way). The luminous
high-$J_{\rm upper}$ emission they see is likely due to the presence
of additional excitation phases in Mrk\,231, either high excitation
PDRs or more likely an X-ray dominated region (XDR) \citep{Spaans08}.
As we see later, we also find evidence for at least two phases of
material in SMM\,J2135.

In comparison to higher-redshift sources, we see that the $^{12}$CO
SLED for SMM\,J2135 has a broadly similar shape to that seen in other
high-redshift SMGs, e.g.\ GN\,20 \,\citep{Carilli10} or SMM\,J16359
\citep{Weiss05b}. Indeed, recent progress with EVLA and GBT has
started to provide $^{12}$CO(1--0) detections of SMGs to complement
the earlier $^{12}$CO(3--2) studies from PdBI
(e.g.\ \citealt{Greve05}).  Using these two transitions, we find that
the $R_{3,1}$ brightness temperature ratio we derive for SMM\,J2135
(Table~4) is comparable to typical SMGs: $R_{3,1}=0.55\pm0.05$ (Ivison
et al.\ 2010c) and $R_{3,1}=0.68\pm0.08$ \citep{Harris10}, suggesting
that the CO SLED results we derive for SMM\,J2135 may be applicable to
the wider SMG population.

To investigate the excitation of the gas reservoir within this galaxy
in more detail we exploit the fact that the shape of the $^{12}$CO SLED
can provide information on the underlying gas density and temperature
distributions and use a spherical large velocity gradient model (LVG;
\citealt{Weiss05b}) to attempt to fit the observed SLED.  LVG
techniques are the most widely used radiative transfer model which can
account for photon transport when spectral lines are optically thick,
and can be used for efficiently solving the radiative transfer
equation when the molecule level populations are not thermalised
(\citealt{Bayet06}).

Thus we next use the LVG code to model the $^{12}$CO SLED.  The model
assumes spherical symmetry and uniform kinetic temperature and
density, a CMB background temperature of T$_{\rm CMB}=2.73 {\rm
  K}(1+z)$ which at $z=2.3$ is $\sim 9$\,K, the collision rates from
\citet{Flower01} with an ortho/para H$_2$ ratio of 3 and a ratio of
the CO abundance to the velocity gradient of
CO\,$/dv/dr=10^{-5}$\,pc\,(km\,s$^{-1}$)$^{-1}$.  To fit absolute line
intensities the model further uses the source solid angle size,
$\Omega_S$, which can be expressed in terms of the equivalent radius
of a face-on disk, $r_0=D_A\sqrt(\Omega_S/\pi)$.  The model returns
the temperature, density and size of the emission region.

Modelling the $^{12}$CO SLED in this way, we find that a single
temperature and density phase is unable to fit the SLED, and that two
or more phases are required.  This is consistent with the implication
of the model of the dust SED and the sub-millimetre morphology: we
need at least two temperature phases to adequately describe this
system. This need for multiple phases to fit the ISM in a
high-redshift galaxy is unusual (however cf. \citealt{Carilli10}),
although it is not unexpected given that such multi-phase ISMs are
commonly required in local galaxies (e.g.\ \citealt{Wild92, Guesten93,
  Aalto95, Mao00, Ward03}).  Thus, motivated by the fit to the dust
SED we identify a two-phase fit to the SLED comprising four clumps and
an extended phase, which provides an adequate fit
(Fig.~\ref{fig:CO_SLED_tot}).  In this model, the low-excitation phase
peaks at $J_{\rm upper}\sim3$ and is diffuse, with
$n\sim10^{2.7}$\,cm$^{-3}$ and T$_{\rm kin}\sim 25$\,K.  The denser,
mildly-excited phase peaks around $J_{\rm upper}=5$--6, has T$_{\rm
  kin}\sim 60$\,K and $n\sim10^{3.6}$\,cm$^{-3}$ and contributes $\sim
60$\% of the total luminosity over all the $^{12}$CO lines
(Fig.~\ref{fig:CO_SLED_tot}).  The low-excitation phase has a
$^{12}$CO SLED which is very similar to the inner disk of the Milky
Way, while the warmer phase is well matched to the inner few hundred
parsecs of the starburst region in NGC\,253.  Within this two phase
model, the total gas mass is
M$_{gas}\sim(4.0\pm0.1)\times10^{10}$\,M$_{\odot}$, a factor of
$\sim3\times$ the mass we determine from $^{12}$CO(1--0), assuming
$\alpha=0.8$, corresponding to an effective $\alpha= 2.0$ (simlar to
that derived for the nucleus of Arp\,220; \citep{Scoville97}). From
here on we adopt this value as the gas mass of our system, but note
some of the differences if we had used the $^{12}$CO line
luminosity and $\alpha=0.8$, as is commonly assumed in the analysis of
high-redshift galaxies.  The combination of this higher value for
$\alpha$ and our estimate of $R_{3,1}$ would increase the gas mass by
a factor of $\sim 4\times$ over that estimated by assuming
$\alpha=0.8$ and $R_{3,1}=1$ as usually assumed at high redshift (see
also \citealt{Ivison10L1L2,Harris10}).  Adopting $\alpha\sim2$ results
in a gas mass fraction in the central 1kpc radius region of SMM\,J2135
of 75\% (and closer to the 100\% for typical SMGs, \citealt{Greve05})
subject to uncertainties on the dynamical mass due to the unknown
configuration of the gas within the system.

\subsubsection{Atomic Carbon}

We have also obtained strong detections of the [C{\sc
    i}]($^3P_1\rightarrow\,^3P_0$) and [C{\sc
    i}]($^3P_2\rightarrow\,^3P_1$) emission lines, and from these we
can begin to investigate the properties of the ISM.  We start by
noting the similar detailed morphologies of the [C{\sc
    i}]($^3P_1\rightarrow\,^3P_0$) and the $^{12}$CO(3--2) and
$^{12}$CO(1--0) lines in Fig.~\ref{fig:all_spec}, which suggests that
the emission is arising from the same mix of phases with
characteristic temperatures of $T\lsim 40$\,K. Indeed observations of
neutral carbon transitions have shown that the ratio of L$'_{\rm
  [CI](^3P_2\rightarrow\,^3P_1)}/$\,L$'_{\rm
  [CI](^3P_1\rightarrow\,^3P_0)}$ provides a sensitive probe of the
temperature of the interstellar medium at moderate densities.  Both
lines have modest critical densities ($n_{\rm crit}\sim
0.3$--$1.1\times10^3$\,cm$^{-3}$) and are therefore often thermalised
in molecular clouds with $n\gsim 10^{3}$\,cm$^{-3}$.  The lines arise
from states with energy levels T$_1=23.6$\,K and T$_2 =62.5$\,K above
the ground state, and thus their ratio is sensitive to the gas
temperature if T$_{\rm gas}\lsim 100$\,K. We derive L$'_{\rm
  [CI](^3P_2\rightarrow\,^3P_1)}/$\,L$'_{\rm
  [CI](^3P_1\rightarrow\,^3P_0)}=0.37\pm0.02$, which is similar to that
measured in nearby starbursts and galactic nuclei such as M\,82 and
NGC\,253 where L$'_{\rm [CI](^3P_2\rightarrow\,^3P_1)}/$\,L$'_{\rm
  [CI](^3P_1\rightarrow\,^3P_0)}\sim 0.33$
\citep{Bennett94,White94,Israel95,Bayet04}, but larger than typically
found in the cooler, dense cores of giant molecular clouds
(e.g.\ \citealt{Zmuidzina88}). To be compatible with previous studies
we define R$_{[\rm CI]}$ as the ratio of the [C{\sc
    i}]($^3P_2\rightarrow\,^3P_1$) to [C{\sc
    i}]($^3P_1\rightarrow\,^3P_0$) temperature integrated line
intensities in K\,km\,s$^{-1}$. In SMM\,J2135 we derive an integrated
line ratio of $R_{[\rm CI]}=0.88\pm0.01$. We can use this ratio to
estimate the excitation temperature T$_{\rm ex}$ following
\citet{Stutzki97}, $T_{\rm ex}=38.8$\,K/$\ln(2.11/R_{[\rm CI]})$, and
derive T$_{\rm ex}=44.3\pm1.0$\,K for SMM\,J2135.

We also follow \citet{Weiss05a} to derive the total mass of neutral carbon via:
\begin{equation}
  M_{\rm CI}=1.902\times10^{-4}Q(T_{\rm ex})e^{T_1/T_{\rm ex}}L'_{\rm [CI](^3P_1\rightarrow\,^3P_0)}
\end{equation}
\smallskip where $Q(T_{\rm ex})=1+3e^{-T_1/T_{\rm ex}}+5e^{-T_2/T_{\rm
    ex}}$ is the [C{\sc i}] partition function and T$_{\rm ex}$ is
defined above.  Using T$_{\rm ex}=44.3\pm1.0$\,K and our measured
luminosity of the upper fine structure line, [C{\sc
    i}]($^3P_1\rightarrow\,^3P_0$), we estimate a carbon mass of
M$_{\rm [CI]}=(9.1\pm0.2)\times10^{6}$\,M$_{\odot}$.  Combining this
with our gas mass estimated from the LVG analysis we derive a [C{\sc
    i}] abundance of M([C{\sc i}])/6M(H$_2$)\,$=(3.8\pm 0.1)\times
10^{-5}$, which is higher than the Galactic value of $2.2\times
10^{-5}$ from \citet{Frerking89}.  Similarly, the ratio of the [C{\sc
    i}] and $^{12}$CO brightness temperatures, L$'_{\rm
  [CI](^3P_1\rightarrow\,^3P_0)}/$\,L$'_{\rm ^{12}CO(1-0)}=0.41\pm
0.02$, falls towards the upper end of the range $0.2\pm 0.2$ observed
in local galaxies \citep{Gerin00,Bayet06}.  A higher abundance of
[C{\sc i}] relative to $^{12}$CO is expected in regions with lower
metallicity \citep{Stark97} where $^{12}$CO is photo-dissociated, in
regions with a high ionisation fraction which may drive the chemistry
to equilibrium at a high [C{\sc i}]/$^{12}$CO ratio, and in starburst
systems with a high cosmic ray flux \citep{Papadopoulos04,Flower94}.

We can also attempt to model the [C{\sc i}] emission within the
framework of our two-phase LVG model.  We find that we are only able
to simultaneously fit the [C{\sc i}] luminosities and ratios if the
carbon abundance is allowed to vary between the two phases.  To fit
the observed R$_{[\rm CI]}$ ratio and the [C{\sc i}] to $^{12}$CO
brightness temperatures the LVG model suggests [C{\sc
    i}/$^{12}$CO]\,$\sim 0.2$ for the warm, dense phase, which is
close to the Galactic value, [C{\sc i}]/$^{12}$CO\,$\sim 0.13$
\citep{Frerking89}, while for the extended, low-density phase we
require an abundance of [C{\sc i}]/$^{12}$CO\,$\sim 3.0$, which
suggests a deficit of oxygen, and hence possibly a very low
metallicity. Using the correlation between [C{\sc i}]/$^{12}$CO and
metallicity (12 + log[O/H]) in \cite{Bolatto00}, for the cool component
we derive $12+$log(O/H)=7.5 and hence a metallicity of
$\sim$Z/Z$\odot$=1/25. This is similar to that of the Sextans dwarf
galaxy, to the lowest metallicity, young starbursts seen locally
\citep{Brown08}, and to the lowest metallicity galaxies found at high
redshift \citep{Yuan09}. However, the high [C{\sc i}]/$^{12}$CO ratio
could also be due to a high cosmic ray flux in the galaxy
\citep{Papadopoulos04,Israel02}.  Indeed a factor of $\sim10\times$
enhancement of cosmic ray flux will yield the high [C{\sc
    i}/$^{12}$CO] ratio we observe and this may not be unlikely given
the high star formation rate density of the star-forming regions
\citep{Swinbank10Nature}. Moreover, an enhanced [C{\sc i}/$^{12}$CO]
ratio has also been observed in the starburst galaxy NGC\,253
\citep{Harrison95,Bradford03} attributed to a high cosmic ray flux.

Finally, the [C{\sc i}] detections can be combined with the
observations of [C{\sc ii}]157.7 from \citet{Ivison10eyelash} to
investigate the ISM cooling.  We can compare the cooling in each of
our lines relative to the bolometric (8--1000$\mu$m) luminosity, to
assess their importance. The fraction of the bolometric luminosity in
the [C{\sc ii}] line is 0.24\%, the rotational $^{12}$CO lines
($^{12}$CO(1--0) to $^{12}$CO(10--9)) contribute 0.09\% and the [C{\sc
    i}] lines result in just 0.03\%.  Therefore, [C{\sc ii}]
dominates the cooling with the total cooling due to $^{12}$CO and
[C{\sc i}] representing $\sim 50$\% of the cooling due to [C{\sc
    ii}]. [C{\sc ii}] is one of the brightest emission lines in
galaxies and can account for 0.1--1\% of the far-infrared luminosity
of the nuclear regions of galaxies (\citealt{Stacey91}); our source
lies well within that range.  This situation is in contrast to
comparably luminous systems in the local Universe, e.g.\ Arp\,220,
where the [C{\sc ii}] emission is just $1.3\times 10^{-4}$ of $L_{\rm
  FIR}$ (Gerin \& Phillips 1998). This most likely results from
saturation of [C{\sc ii}] in very high-density star-forming regions
(Luhman et al.\ 1998). The proportionally stronger [C{\sc ii}]
emission we see in SMM\,J2135 may then reflect slightly lower
densities and more extended star formation (Ivison et al.\ 2010c), or
be due to the lower metallicity of the gas compared to local ULIRGs
(\citealt{Israel96}; \citealt{Maiolino09}).

Overall, our observations of atomic carbon suggest that it arises from
a cool phase within the galaxy and that integrated over the whole
galaxy the [C{\sc i}]/$^{12}$CO ratio is higher than the Galactic
value. However, we also find evidence for variation in the ratio
within the system, with the cool, low-density phase having
significantly enhanced [C{\sc i}], compared to $^{12}$CO, suggesting
lower metallicity in this phase or a higher cosmic ray flux in the
overall system.

\subsubsection{HCN(3--2)}
Our spectra also cover emission lines from a number of species other
than $^{12}$CO; notably, we detect HCN(3--2) emission (Table~2). HCN
is an effective tracer of dense gas due to its high dipole moment,
requiring $\sim100 \times$ higher densities for collisional excitation
than $^{12}$CO(1--0). Indeed, HCN is one of the most abundant
molecules at densities $n\gsim 3\times10^{4}$\,cm$^{-3}$ (compared to
critical densities of $\gsim 500$\,cm$^{-3}$ for low-$J_{\rm upper}$
levels of $^{12}$CO). We note that the velocity centroid of the HCN
emission is redshifted by approximately $+230\pm 100$\,km\,s$^{-1}$,
relative to the nominal systemic redshift of the system derived from
$^{12}$CO.  As we discuss below, this hints that the HCN emission may
arise predominantly from only one of the kinematic components within
the galaxy.

%
% Figure 3
%
\begin{figure}
\centerline{
  \psfig{figure=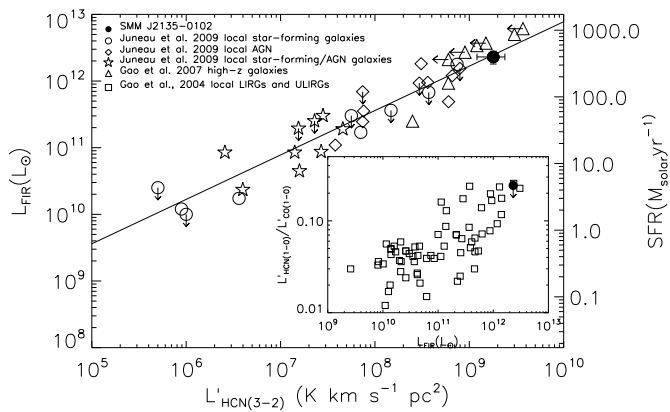,width=3.5in,angle=90}}
\caption{The relation between far-infrared luminosity and HCN
  luminosity for galaxies.  We compare our observations of SMM\,J2135
  with the correlation seen locally and find that this high-redshift
  galaxy follows the tight correlation seen in local galaxies
  (\citealt{Juneau09}). In the main panel we plot observations from a
  sample of 34 local galaxies (star-forming galaxies, star-forming
  galaxies with AGN and strong AGN) taken from \citet{Juneau09}. In the
  inset panel we compare our upper limit on the ratio of L$'_{\rm
    HCN(1-0)}$\,/\,L$'_{\rm ^{12}CO(1-0)}$ to the ratio in local LIRGs
  and ULIRGs from \citet{Gao04b} showing that this upper limit is
  consistent with the \citet{Gao04b} relation.  Again we see that the
  properties of the dense gas and far-infrared emission in this galaxy
  are similar to those seen locally.}
\label{fig:gao04}
\end{figure}

It has been shown that the HCN line luminosity is tightly correlated
with far-infrared luminosity in local spirals and ULIRGs, with a
linear relation holding over at least three decades in luminosity, and
a similar relation may hold at high redshift (Fig.~\ref{fig:gao04};
\citealt{Gao04b,Gao07}).  This suggests that HCN is a good tracer of
the dense gas which fuels the massive star formation, which in turn is
responsible for the far-infrared emission. In Fig.~\ref{fig:gao04} we
show the relation between far-infrared luminosity and HCN(3--2) line
luminosity for local spirals, LIRGs and ULIRGs
(\citealt{Gao04b,Juneau09}).  SMM\,J2135 lies at the high-luminosity
end of this correlation, corresponding to the highest-luminosity local
ULIRGs, and is consistent with the relation.  In the inset panel on
Fig.~\ref{fig:gao04} we also compare our source to local LIRGs and
ULIRGs in \citet{Gao04b}.  We show the upper limit on L$'_{\rm
  HCN(1-0)}$\,/\,L$'_{\rm ^{12}CO(1-0)}$ using our 3-$\sigma$ upper
limit for HCN(1--0) flux, demonstrating that this is also consistent
with the increasing L$'_{\rm HCN}$\,/\,L$'_{\rm ^{12}CO}$ ratio seen
with increasing L$_{\rm FIR}$ at low redshift.

It is also possible to estimate a total gas mass from the HCN
luminosity \citep{Gao05}.  Using our 3-$\sigma$ upper limit on the
HCN(1--0) luminosity we derive a limit on the dense gas mass of
M$_{\rm dense}$(H$_2$)\,$=\alpha_{\rm HCN(1-0)}$\,L$'_{\rm HCN} \ll
4.5 \times 10^{10}$\,M$_{\odot}$ with $\alpha_{\rm HCN} \ll 10$
\citep{Gao04b}. We caution that there is at least a factor $3\times$
uncertainty arising from the uncertainty in $\alpha_{\rm HCN}$.  The
value of $\alpha_{\rm HCN}$ we adopt is appropriate for a virialised
cloud core; however, ULIRGs and LIRGs usually have a much higher HCN
brightness temperature, resulting in their $\alpha_{\rm HCN}$ being
much lower \citep{Gao04b}, and hence this is reported as an upper limit
on the gas mass.  Since this gas mass is consistent with the mass
derived from our LVG analysis of $^{12}$CO, this suggests that with
better estimates of $\alpha_{\rm HCN}$, HCN may be a promising route
to constraining the dense gas masses of high-redshift galaxies in the
future. However, we note that the consistency of $^{12}$CO-based and
HCN-based gas mass estimates may be coincidental if these lines
trace different phases. We further caution that at least two HCN lines
may be required to derive the correct value of $\alpha_{\rm HCN}$ due
to the strong variations in HCN(4--3)/HCN(1--0) and
HCN(3--2)/HCN(1--0) luminosity ratios locally \citep{PPP06}.

As with [C{\sc i}], we also attempted to use our two-phase LVG model
from the $^{12}$CO SLED to model the HCN(3--2) and HCN(1--0) emission
from this system. However, we find that any models which fit the HCN
SLED, in particular the bright HCN(3--2) emission, over-predict the
luminosities of the $^{12}$CO SLED and result in gas masses which are
greater than the dynamical mass. A similar problem has been identified
in local ULIRGs which exhibit enhanced HCN line emission and high
ratios of L$'_{\rm HCN}$\,/\,L$'_{\rm ^{12}CO}$ due to a higher
proportion of dense gas in these strong starburst systems
\citep{SolomonVandenBout05,Gao07}.

\subsubsection{H$_2$O}
Finally we comment on the limits obtained for H$_2$O. H$_2$O emission
arises from the warm, dense gas in the densest regions of a starburst
or around an AGN.  The strength of the emission is thus a probe of the
radiation density and so reflects the compactness of the far-infrared
source.  The H$_2$O\,(2$_{1,1}$--2$_{0,2}$) (731.681\,GHz) emission
line has recently been detected in the local ULIRGs Mrk\,231 and
Arp\,220 (e.g. \citealt{GonAlf10}) revealing warm, dense material,
possibly arising from an XDR associated with AGN activity (see also
\citealt{Spaans08} and \citealt{VdW10}).  We compare our limits on the
H$_2$O line luminosities (as a fraction of bolometric luminosity) with
Mrk\,231 and Arp\,220, where the luminosity ratios are L$_{\rm
  H_2O}$/L$_{\rm FIR}\sim1$--$2\times10^{-6}$.  We derive limits on
the line luminosity ratio of L$_{\rm H_2O}$/L$_{\rm
  FIR}<1$--$5\times10^{-6}$ from the H$_2$O\,(5$_{1,5}$--4$_{2,2}$)
and H$_2$O\,(2$_{1,1}$--2$_{0,2}$) transitions.  These suggest that if
an XDR exists within SMM\,J2135 it is unlikely to be more luminous
than those seen in these local AGN-dominated ULIRGs.  This conclusion
is reinforced by the relatively low luminosity of the high-$J_{\rm
  upper}$ $^{12}$CO lines in SMM\,J2135 compared to the substantially
brighter high-$J$ lines in Arp\,220 and Mrk\,231 which again require
an XDR to fit their $^{12}$CO SLEDs \citep{Spaans08,VdW10}.

\subsection{Integrated Properties: Physical Properties of the ISM}
\label{sec:PDRs}

When massive stars are formed around a molecular cloud, their
ultra-violet (UV) radiation changes the chemical properties of the
cloud's surface layers; the far-UV flux photodissociates the outer
layer and ionises the material (such as carbon).  This layer then
cools primarily via atomic fine structure lines of [O{\sc i}], [C{\sc
    ii}], [C{\sc i}], and the rotational $^{12}$CO lines.  Emission
from these different coolants arises from different depths within the
PDR \citep{Kramer04}, such that the surface layers are dominated by
emission from hydrogen, [C{\sc ii}] and oxygen, although further into
the star-forming clouds (as the extinction, A$_V$, increases),
hydrogen becomes molecular, whilst the ionisation of carbon declines,
with [C{\sc ii}] becoming [C{\sc i}] and then combining into CO.
Moreover, with increasing density, the higher-$J_{\rm upper}$
$^{12}$CO emission becomes stronger, as do other molecular gas tracers
such as HCN, CS and CN \citep{Kaufman99}. Deeper into the cloud, the
gas is molecular, but still has a higher temperature than in the
far-UV shielded core.  Significant effort has been devoted to
modelling (and predicting) the fine structure and molecular emission
line ratios associated with these star-forming regions using PDR
models.  These account for variations in density, temperature,
clumpiness and time-dependent chemistry (e.g.\ \citealt{Meijerink07,
  Kaufman99}) and using their predictions of the ratios of [C{\sc
    ii}], [C{\sc i}] and $^{12}$CO we can investigate the typical
far-UV intensity and ISM density within the star-forming regions in
SMM\,J2135.

\subsubsection{PDR models}
\label{sec:PDRmodels}

To model the full-range of observed line ratios from SMM\,J2135 we use
the PDR models of \citet{Kaufman99} and \citet{Meijerink07} in which
the emission is determined by the atomic gas density ($n$; the density
of H nuclei) and the incident far-UV radiation field from massive
stars ($h\nu=6$--13.6\,eV) which is expressed in terms of $G_0$ (where
$G_0$ is the average far-UV radiation field in the Milky Way and is
expressed in Habing units:
$1.6\times10^{-3}$\,erg\,s$^{-1}$\,cm$^{-2}$).  Simple PDR models
(including one of the two models we compare to; \citealt{Kaufman99})
calculate the line emission generated by a cloud illuminated only on
one side. However, for observations of external galaxies it is not
clear that such a ``single cloud'' model will adequately describe the
observed molecular lines. For clouds illuminated on all sides we note
that an observer will detect optically thin radiation emitted from
both the near and far sides of the cloud, but only see optically thick
emission from the near side. Generally, low-$J_{\rm upper}$ $^{12}$CO
transitions are assumed to be optically thick for the nominal
$A_{V}=10$ cloud depths in the PDR models \citep{Hailey10}, whilst
[C{\sc ii}], [C{\sc i}], $^{13}$CO and L$_{\rm FIR}$ are assumed to be
optically thin (although note our earlier estimates of the high column
densities and extinction in this system).  We have not applied any
corrections to the emission lines to account for these differences
between optically thick and optically thin emission.  Instead, we also
compare to more complicated models involving two or more clouds, with
differing densities and incident radiation fields which have been
constructed by \cite{Meijerink07}. The implementation of these models
assumes all emission lines are coming from the same gas in the same
region, which we know to be a vast over-simplification; however, these
solutions do provide us with order of magnitude estimates of the
characteristic conditions in the system.

%
% Figure 4
%
\begin{figure*}
\centerline{\psfig{figure=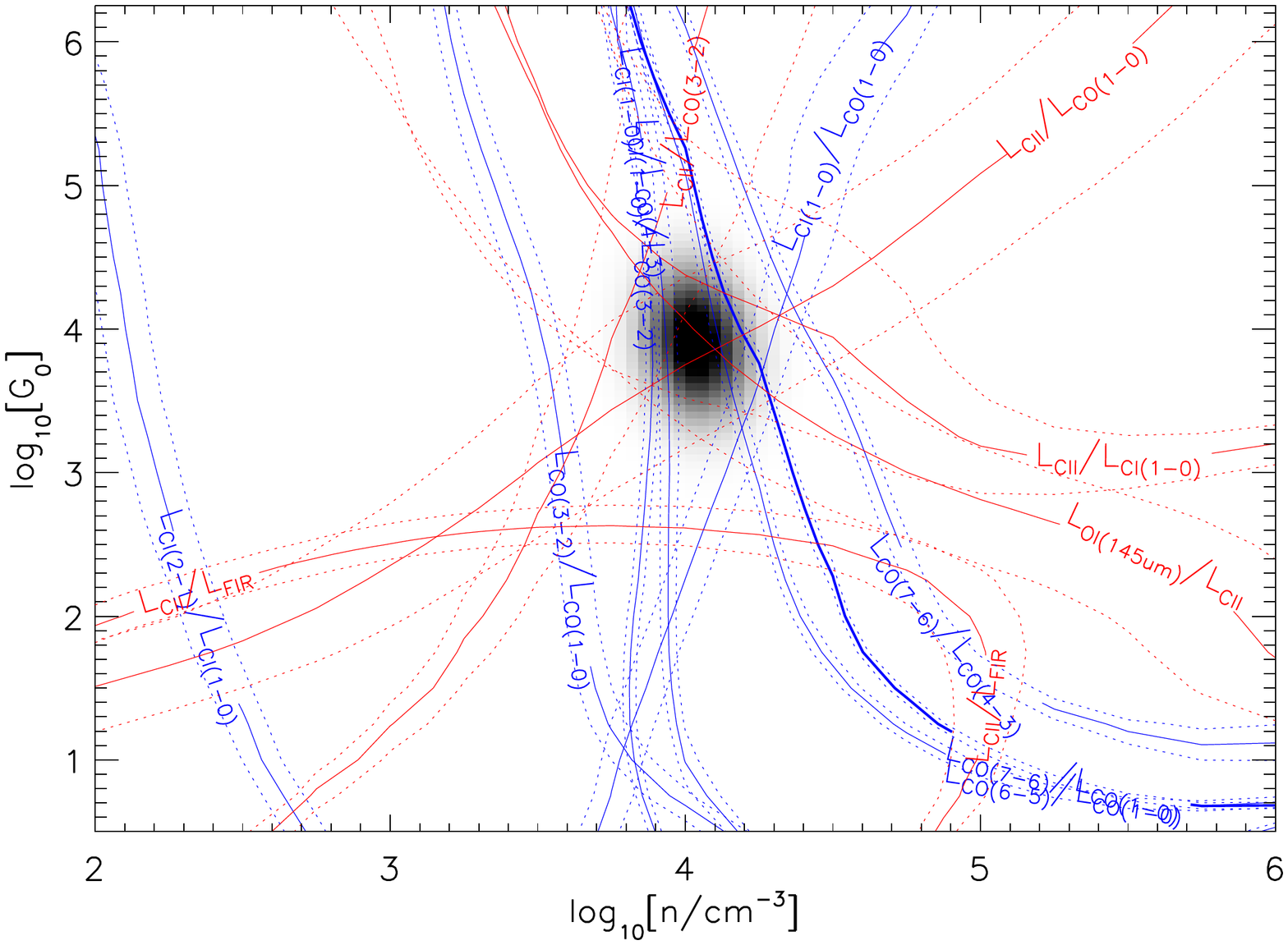,width=3.7in,height=2.7in,angle=90}
\psfig{figure=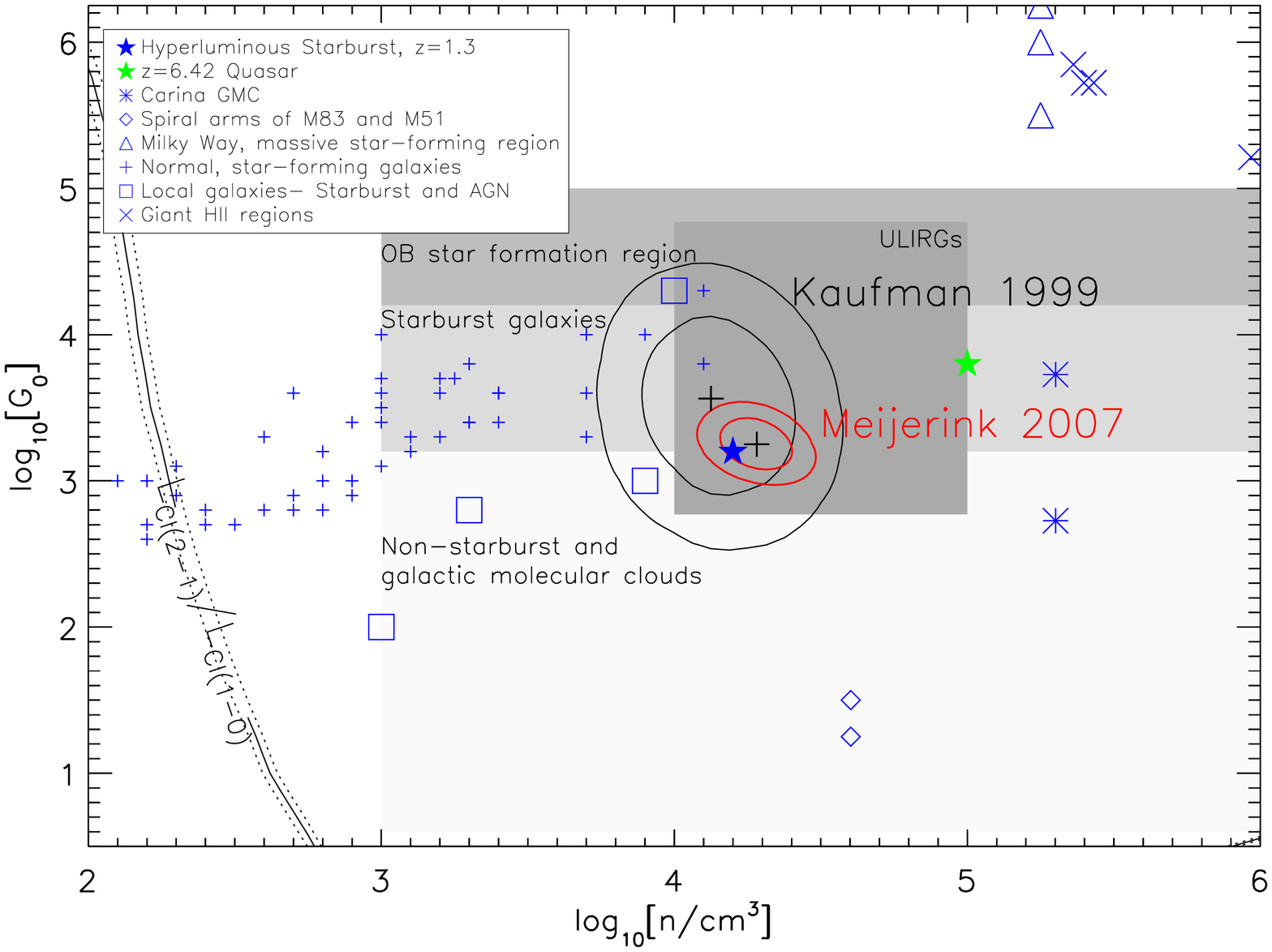,width=3.5in,height=2.5in,angle=90,clip=t}}

\caption{{\it Left:} Luminosity line ratios (in units of L$_{\odot}$)
  from $^{12}$CO, [C{\sc i}] and [C{\sc ii}] as a function of density
  and far-UV flux ($G_{0}$ in units of Habing field), from the PDR
  models of \citet{Kaufman99}.  Tracks are drawn for the measured
  ratios in solid lines with the 1-$\sigma$ errors as dotted lines.
  The line ratios within SMM\,J2135 intercept at $n\sim
  10^{4.1}$\,cm$^{-3}$ and $G_{0}\sim10^{3.6}$ Habing fields and
  $n\sim 10^{4.3}$\,cm$^{-3}$ and $G_{0}\sim10^{3.3}$ Habing fields
  for the \citet{Kaufman99} and \citet{Meijerink07} models
  respectively. To better display the preferred regions of parameter
  space we combine the probability distributions from all of these
  lines (excluding $L_{\rm [CI](2-1)}/L_{\rm [CI](1-0)}$ for reasons
  explained in \S~\ref{sec:PDRs}) to derive a peak likelihood
  solution for each model. {\it Right}: The same parameter space for
  the models, now comparing the peak likelihood solutions from the
  model grids to the derived values of $G_{0}$ and density ($n$) for
  various low- and high-redshift galaxies and molecular clouds.  We
  also indicate the regions of parameter space which are typically
  encompassed by Galactic OB star-formation regions, starburst
  galaxies and non-starburst and Galactic molecular clouds from
  \citet{Stacey91}.  This shows that the line ratios for SMM\,J2135
  are consistent with those typically found in local starbursts or
  ULIRGs. The local galaxy that lies within the contours of the peak
  likelihood for the Kaufman models is NGC\,253, a spiral starburst
  galaxy \citep{Negishi01}. We also show the peak likelihood solution
  derived from the PDR models of \citet{Meijerink07} which provide a
  concomitant solution. The contours represent the 1- and 2-$\sigma$
  limits compared to the best-fit solution. The peak likelihood
  solutions are derived without including the L$_{\rm
   [CI](2-1)}$\,/\,L$_{\rm [CI](1-0)}$ track, which we plot on both panels
  to show its discrepancy from the general solution from the other
  lines (see \S~\ref{sec:PDRs}).}
\label{fig:cont_diag}
\end{figure*}

In Fig.~\ref{fig:cont_diag} we show the model grid of $n$ and $G_0$
values for various emission line ratios for SMM\,J2135. The $^{12}$CO
$J/(J-1)$ and [C{\sc i}]/$^{12}$CO line luminosity ratios vary over
six orders of magnitude in $G_0$; however, the [C{\sc ii}]/FIR and
[C{\sc ii}]/$^{12}$CO ratios appear to break this degeneracy.  Indeed,
in the $n\sim10^{3}$--$10^{5}$\,cm$^{-3}$ regime, these provide a
strong constraint on G$_0$.  Using the two different PDR models we
convert the allowed parameter space for each line ratio to derive the
peak likelihood solution, which is shown in Fig.~\ref{fig:cont_diag}
for both models.  We find that the solutions cluster around moderate
densities ($n\sim 10^{4}$\,cm$^{-2}$), similar to those claimed for
local ULIRGs from PDR modelling of H$_2$ emission \citep{Davies03}.
The peak likelihood solutions use all the line ratios available aside
from the L$'_{\rm [CI](^3P_2\rightarrow\,^3P_1)}/$\,L$'_{\rm
  [CI](^3P_1\rightarrow\,^3P_0)}$ ratio. This line's track deviates
strongly from the preferred solution from the other lines. We expect
this reflects the sensitivity of this line ratio to details of the PDR,
such as geometry, and we discuss specific problems with the PDR model treatment
of [C{\sc i}] below.

The best-fit solution to our suite of line ratios is
$n=10^{4.1\pm0.3}$cm$^{-3}$ and $G_{0}=10^{3.6\pm0.7}$ Habing fields
for the PDR models from \citet{Kaufman99} and $n\sim 10^{4.3\pm
  0.2}$\,cm$^{-3}$ and $G_0\sim 10^{3.3\pm 0.2}$ Habing fields from
\citet{Meijerink07}.  The best-fit solutions are within 1-$\sigma$ and
are in agreement with earlier estimates based on just $^{12}$CO(1--0),
[C{\sc ii}] and L$_{\rm FIR}$ in \citet{Ivison10eyelash}.  As we see
from Fig.~\ref{fig:cont_diag}, these physical conditions are similar
to local starburst galaxies and ULIRGs \citep{Stacey91,Davies03} and
at the upper end of both density and $G_{0}$ for normal star-forming
galaxies \citep{Malhotra01}, for example the radiation field is
$\sim1000\times$ more intense than in the disk of the Milky Way. The
characteristic density compares well to the dense star-forming cores
of Galactic molecular clouds, or to the central compact nuclear gas
disks of local ULIRGs such as Arp\,220 \citep{Downes98, Sakamoto99}
and is also consistent with the integrated $^{12}$CO SLED (see
\S\ref{sec:cosled}).  This characteristic density we derive is greater
than the critical density for [C{\sc ii}], [C{\sc i}] and CO(1--0),
above which the intensity of the lines saturates. Similarly, the
G$_{0}$ value we derive suggests the conditions in the PDRs are at the
limit of the steady state solution \citep{Kaufman99}; with radiation
pressure on grains comparable to turbulence, suggesting that the PDRs
may be ionising material.

Our results imply that the underlying physics of the star formation
occurring in the system is similar to the dense star-forming cores of
Galactic molecular clouds (characteristic of massive OB star-forming
regions), even though the total energetics of the system is far more
extreme than in the disk of our galaxy.  Interestingly, \citet{Tey10}
have recently used hydrodynamic simulations of major mergers of disk
galaxies to show that the density of gas at which a majority of the
star formation in these systems occurs is n$\sim10^{4}$\,cm$^{-3}$, as
we find here.

Of all the line ratios we analyse, the most obvious outlier in the PDR
modelling is the [C{\sc i}]($^3P_2\rightarrow\,^3P_1$)/[C{\sc
    i}]($^3P_1\rightarrow\,^3P_0$) emission line ratio
(Fig.~\ref{fig:cont_diag}).  However, we note that this line ratio is
one of the most sensitive to $G_0$ and $n$ and it would only require a
modest increase in the predicted [C{\sc i}]($^3P_1\rightarrow\,^3P_0$)
flux in the models, $\sim 3\times$ stronger, to shift this track into
agreement with the solution based on the bulk of the other ratios at
$n\sim 10^{4}$\,cm$^{-3}$ and $G_0\sim 10^{3.5}$.  This problem may
arise because the PDR models assume a particular geometry, which
determines the ratio of diffuse to dense gas, and assume a homogeneous
medium (an unrealistic assumption for our source).  The resulting
luminosity ratios based on $^{12}$CO/[C{\sc i}] and [C{\sc i}] are
sensitive to this geometry and the [C{\sc
    i}]($^3P_2\rightarrow\,^3P_1$)/[C{\sc
    i}]($^3P_1\rightarrow\,^3P_0$) ratio is particularly sensitive to
the homogeneity of the medium \citep{Spaans96}, as demonstrated by the
difficulty the models have in reproducing the [C{\sc
    i}]($^3P_1\rightarrow\,^3P_0$) or [C{\sc
    i}]($^3P_2\rightarrow\,^3P_1$) emission lines even in tranquil
environments (e.g.\ \citealt{Pin07}; \citealt{PPP04b}).  Indeed,
\citet{PPP04b} argue that [C{\sc i}] is distributed relatively
ubiquitously in molecular clouds (the critical density for excitation
of both [C{\sc i}]($^3P_1\rightarrow\,^3P_0$) and [C{\sc
    i}]($^3P_2\rightarrow\,^3P_1$) is roughly that of $^{12}$CO(1--0);
\citealt{SolomonVandenBout05}). This is contrary to PDR theory, which
places [C{\sc i}] in a narrow [C{\sc ii}]/[C{\sc i}]/$^{12}$CO
transition zone on the surface of far-UV illuminated molecular clouds,
and potentially makes [C{\sc i}] a very effective tracer of H$_2$
mass.  Therefore our low ratio may be arising from geometrical optical
depth effects in which the [C{\sc i}] effectively follows the volume
(rather than the PDR surface), which would naturally lower the [C{\sc
    i}]($^3P_2\rightarrow\,^3P_1$) /[C{\sc
    i}]($^3P_1\rightarrow\,^3P_0$) ratio (e.g.\ see
\citealt{Kramer08}) as would happen from any process which enhances
        [C{\sc i}] emission through the cloud's volume (e.g. cosmic
        rays).  We also caution that the [C{\sc
            i}]($^3P_2\rightarrow\,^3P_1$) /[C{\sc
            i}]($^3P_1\rightarrow\,^3P_0$) ratio has been measured in
        only a few local galaxies (e.g.\ \citealt{Bayet06}) and so the
        fine structure excitation of neutral carbon has yet to be
        fully characterised and calibrated locally.

\subsubsection{Limits on $^{13}$CO, CN, HNC and HCO$^+$}

Our observations have also simultaneously covered a number of dense
gas tracers and from these we can place limits on the line
luminosities of $^{13}$CO, CN, HNC and HCO$^+$. We can then ask if
these limits are consistent with our preferred characteristic density
and UV radiation field from the PDR model.

First we consider the limits on the $^{13}$CO.  The $^{13}$C nuclei
are synthesised by CNO processing of $^{12}$C nuclei from earlier
stellar generations ($^{12}$C is produced from He burning on rapid
timescales in massive stars).  Since $^{13}$C is produced more slowly
in intermediate mass stars during the red giant phase (at
$\sim1$\,Gyr), the $^{12}$C/$^{13}$C ratio has been proposed as a
diagnostic of the nucleosynthesis history in galaxies. A high ratio
would then be seen at high redshift as the young starbursts will have
had insufficient time to form secondary nuclei of $^{13}$CO.  In the
Milky Way and nearby star-forming galaxies, the $^{12}$CO/$^{13}$CO
line luminosity ratio (L$'_{^{12}\rm CO}$/L$'_{^{13}\rm CO}$) is
$\sim5$--10.  However, local ULIRGs have shown a deficiency of
$^{13}$CO (possibly due to gas inflow; Rupke et al.\ 2008), which
raises the $^{12}$CO/$^{13}$CO line ratio to $\sim30$ (Greve et
al. 2009).  Indeed, recently, Henkel et al.\ (2010) showed that the
$^{12}$CO/$^{13}$CO luminosity line ratio in the Cloverleaf quasar
($z$=2.6) is L$'_{^{12}\rm CO}$/L$'_{^{13}\rm CO}$=40$\pm$17,
suggestive of a deficiency of $^{13}$CO.  Within SMM\,J2135, we
estimate limits on the ratio of $>$5.8 and $>$6.5 for the $J_{\rm
  upper}$=1 and $J_{\rm upper}$=3 transitions respectively.  These
limits are consistent with the Milky Way, but more sensitive
observations are required to place firm conclusions as to whether a
deficiency exists similar to that observed in other local ULIRGs and
high-redshift galaxies.

We can compare these limits to the predictions from the PDR models of
Meijerink et al.\ (2007). The model predictions are L$_{^{12}\rm
  CO}$/L$_{^{13}\rm CO}$=2.13 and 2.08 for the $J_{\rm upper}$=1 and
$J_{\rm upper}$=3 transitions respectively, indicating lower $^{13}$CO
abundance in SMM\,J2135-0102 than the model predicts. This may arise
due to a number of effects.  It is known that the line luminosity
ratios of giant molecular clouds are lower than galaxy-integrated
properties due to the lower opacity of $^{13}$CO compared to $^{12}$CO
in the inter-cloud medium. Since we are measuring galaxy integrated
properties, this would naturally lower the ratio.  Clearly it would be
useful to obtain a more sensitive observation of the $^{13}$CO to
determine the extent of this discrepancy with PDRs.

Finally, we note that the line luminosity ratio of the dense gas
tracers of CS(2--1)/$^{12}$CO(1--0)$<$0.13 is lower than that
predicted from the PDR model (0.43). In contrast, the
HNC(3--2)/$^{12}$CO(1--0)$<1.3$ and
HCO$^+$(3--2)/$^{12}$CO(1--0)$<1.2$ are consistent with those
predicted by the PDR models (1.3 and 0.15 respectively).  A more
detailed test of the model prediction for these dense gas tracers will
again require more sensitive observations.

\subsection{Kinematically Resolved Properties: Decomposed $^{12}$CO SLED}
\label{sec:dec_cosled}

%
% Figure 5
%
\begin{figure}
\centerline{
  \psfig{figure=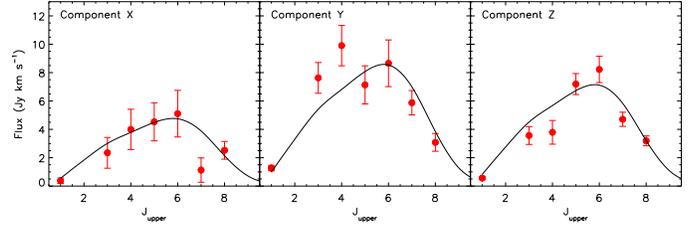,width=3.5in,angle=90}}
\caption{The $^{12}$CO SLEDs for the three kinematic components seen within the
  $^{12}$CO spectra (from left-to-right for the components at X, Y and
  Z at $\sim-180$, 0 and +200\,km\,s$^{-1}$ respectively).  We
  overplot on each of these a scaled version of the best-fit LVG model
  from the combined SLED.  This shows that two components, {\it X} and
  {\it Z}, also peak at around $J_{\rm upper}\sim 6$, while component
  {\it Y} appears to peak slightly lower, $J_{\rm upper}\sim4$,
  suggesting that there may also be temperature differences between the
  kinematic sub-components within the galaxy.  Interestingly,
  component {\it Y} also appears to contain the bulk of the cold gas
  as traced by $^{12}$CO(1--0) compared to either {\it X} or {\it
    Z}. }
\label{fig:CO_SLED_tot2}
\end{figure}

%
% Figure 6
%
\begin{figure}
\centerline{
\psfig{figure=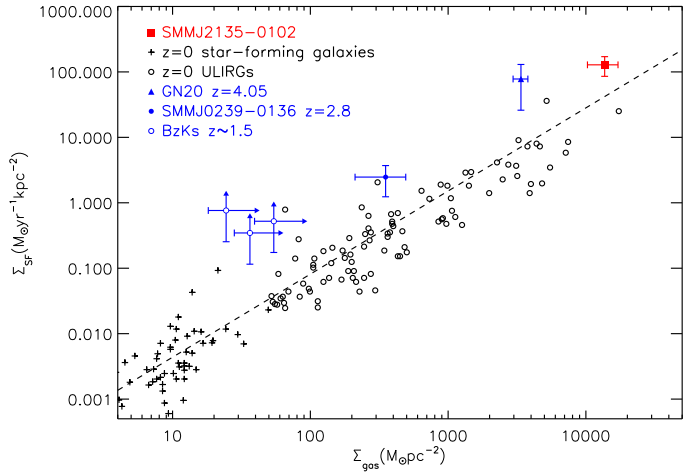,width=3.6in,angle=90}}
\caption{The correlation between measured gas and SFR densities for
  SMM\,J2135 compared to samples of local star-forming galaxies and
  ULIRGS from \citet{KS98} and high-redshift, as derived from
  $^{12}$CO(1--0).  The dashed line is the best fit power law to local
  data with $\Sigma_{\rm SF}\propto\Sigma_{\rm gas}^{1.27}$ (Genzel et
  al. 2010).  We see that the star formation within SMM\,J2135 is
  consistent with this relation assuming $r_{\rm gas}=r_{\rm
    SF}=1$\,kpc. SMM\,J02399$-$0136 is taken from Ivison et
  al.\ (2010), GN20 from \citealt{Carilli10} and the BzKs from
  \citet{Aravena10} which are unresolved in $^{12}$CO(1--0), hence the
  arrows representing the lower limits in measured gas density.}
\label{fig:KS}
\end{figure}

%
% Table 4
%
\begin{table*}
\begin{center}
\small {\centerline{\sc Table 4: Line ratios} 
\smallskip
\begin{tabular}{lccccccccc}
\hline
Component      & M$_{\rm H_2+He}$ & $R_{3,1}^a$      & $R_{4,1}$      & $R_{5,1}$        & $R_{6,1}$     & $R_{7,1}$        & $R_{8,1}$       & $\frac{L'_{\rm CI(2-1)}}{L'_{\rm CI(1-0)}}$ & $\frac{L'_{\rm CI(1-0)}}{L'_{\rm CO(1-0)}}$\\
               & ($10^{9}$\,M$_{\odot}$) &          &               &                 &              &                  &                &                                  & \\
\hline 
Total          & $40.0\pm0.3$    & $0.68\pm 0.03$ & $0.50\pm 0.04$ & $0.35\pm 0.02$ & $0.28\pm0.02$ & $0.119\pm0.008$ & $0.064\pm0.005$ & $0.37\pm 0.02$  &  $0.41\pm0.02$ \\
{\it Z }       & ~$10.1\pm1.8$    & $0.71\pm 0.19$ & $0.42\pm 0.12$ & $0.51\pm 0.11$ & $0.41\pm0.09$ & $0.17\pm 0.04$  & $0.09\pm 0.01$  & $0.85\pm 0.21$  & $0.32\pm0.10$ \\
{\it Y}        & ~$23.1\pm3.2$   & $0.67\pm 0.14$ & $0.49\pm 0.10$ & $0.22\pm 0.05$ & $0.19\pm0.05$ & $0.09\pm0.02$   & $0.04\pm0.01$   & $0.19\pm 0.05$ & $0.43\pm0.09$ \\
{\it X}        & ~$6.8\pm3.2$    & $0.70\pm 0.48$ & $0.67\pm 0.42$ & $0.48\pm 0.29$ & $0.38\pm0.23$ & $0.06\pm0.06$   & $0.11\pm0.06$   & $0.43\pm 0.27$ & $0.38\pm0.28$ \\
\hline
\label{tab:rfac}
\end{tabular} }
\end{center}

\footnotesize{Masses and line luminosity ratios for individual kinematic components. \\ $^aR_{J,1}$ represents L$'_{\rm CO(J+1-J)}$/L$'_{\rm CO(1-0)}$}

\end{table*}

%Table 3
\begin{table}
\begin{center}
\small \centerline{\sc Table 3.}  \centerline{\sc Model Kinematic 
  Parameters}
\vspace{0.2cm}
\begin{tabular}{lcc}
\hline
Component & $v$ & $\sigma$  \\ %\hline
&  (km s$^{-1}$) &  (km s$^{-1}$) \\ %\hline
\hline

{\it Z$_2$}            & $-167 \pm 9$ & $75 \pm 8$ \\
{\it Z$_1$}            & $28 \pm 9$   & $75 \pm 8$ \\
{\it Y  }              & $165 \pm 13$ & $157 \pm 17$ \\
{\it X  }              & $396 \pm 9$  &  $76 \pm 9$   \\
\hline
\label{tab:decomp}
\end{tabular}
\end{center}
\footnotesize{Notes: Velocities and line widths of the three kinematic
  components seen in the CO and [C{\sc i}] spectra.  The velocities are
  given with respect to a heliocentric redshift of $z=2.32591$.}
\end{table}

With the high signal-to-noise of our observations of $^{12}$CO and
[C{\sc i}] in Fig.~\ref{fig:all_spec}, it is obvious that the spectra
show significant kinematic structure and that more intriguingly this
structure appears to vary between the different transitions and
species.  We therefore attempt to fit a model to the $^{12}$CO spectra
and extract $^{12}$CO SLEDs for each kinematic component, in order to
search for excitation structure within the system.

To construct a kinematic model for the $^{12}$CO emission we fit the
composite spectrum (Fig.~\ref{fig:all_spec}) with a series of
increasingly complex models.  We find that a double Gaussian profile
fit provides a $\sim 14$-$\sigma$ improvement over a single Gaussian
profile, with a triple profile giving a further $\sim6$-$\sigma$
improvement over the double Gaussian profile.  Although this model
reasonably describes the line profile of the system, P. Cox et al. (in
prep) use the velocity structure from high-resolution interferometry
with PdBI up to $J_{\rm upper}=7$ to derive a map of the dynamics of
SMM\,J2135 on $\sim 100$-pc scales. The system can be described by two
interacting disks: the lower-redshift of these displays a
double-horned profile (termed {\it Z$_1$} and {\it Z$_2$}), with a
second, {\it Y}, at higher velocity which appears to connect to a
further structure, {\it X}, extending to much higher velocity. Based
on this interpretation we construct a model which comprises three
components: two single Gaussians representing {\it X} and {\it Y} and
a double-Gaussian for {\it Z} (constrained to have equal line widths
and intensities for the two components).  We fit this model to the
composite spectrum, allowing the intensity, width and velocity of each
component to vary and show the resulting best fit to the average line
profile in Fig.~\ref{fig:all_spec} and report its parameters in
Table~3. This fit has a $\chi^2$ of $\sim 20$ for seven degrees of
freedom and provides an adequate description of the composite line
profile.

Using the best-fit model parameters from the composite $^{12}$CO
spectrum, we apply the fit to each $^{12}$CO spectrum in turn, in
order to extract the SLEDs for the three components in the model.
Beginning with the average model parameters, we perturb the best fit
randomly by up to $\pm20$km\,s$^{-1}$ and fit this new realisation to
each of the $^{12}$CO spectra in turn, allowing only the relative
intensities of the components to vary.  We then sum the $\chi^2$ for
the fit to each line for each model and determine the parameters for
which the total $\chi^2$ is a minimum.  We show the best fit model's
line profile on each line in Fig.~\ref{fig:all_spec} and report their
parameters in Table~3. To calculate errors, we allow the best fit
parameters of the kinematic model to randomly vary (by their
1-$\sigma$ uncertainties) and run 10$^5$ Monte Carlo simulations,
determining the 1-$\sigma$ variance in the range of fluxes from those
models where the total $\chi^2$ of the model fit is within $\Delta
\chi^2=21$ of the best fit (since we have seven emission lines and fit
with three free parameters to each line). The errors on the parameters
for the model fit to the average (velocity and dispersion in Table~3)
are calculated using the standard deviation of the parameters for all
models for which the total $\chi^2$ of the model fit is within $\Delta
\chi^2=21$ of the best fit; these are quoted in Table~3.

The $^{12}$CO SLEDs for the three individual components are shown in
Fig.~\ref{fig:CO_SLED_tot2}.  We derive the gas masses for each
component using the $^{12}$CO(1--0) emission line and assuming our
effective conversion factor $\alpha=2.0$ and list these in Table~4.
The most highly excited components are {\it X} and {\it Z}, peaking at
$J_{\rm upper}\sim 6$, and containing $\sim15$\% and $\sim 25$\% of
total gas mass respectively (Table~4).  Component {\it Y}, which we
associate with the core of the second interacting system, is somewhat
less excited, peaking at $J_{\rm upper}\sim4$, and contains most of
the gas in the system ($\sim60$\% of the total gas mass and a majority
of the cold gas).

For each of the SLEDs, we perform further LVG modelling as applied to
the combined spectrum.  However, within the errors, we cannot measure
any temperature difference between the three components although there
is a hint that {\it Y} is cooler than {\it X} or {\it Z}. However,
while the [C{\sc i}]($^3P_2\rightarrow\,^3P_1$) and [C{\sc
    i}]($^3P_1\rightarrow\,^3P_0$) lines have lower signal-to-noise
than the $^{12}$CO lines, we can also decompose them using the same
method. Table~4 shows the ratio of L$'_{\rm
  [CI](^3P_2\rightarrow\,^3P_1)}/$\,L$'_{\rm
  [CI](^3P_1\rightarrow\,^3P_0)}$ for the different components, with
{\it Z} having the largest ratio and {\it Y} having the smallest. The
corresponding excitation temperatures are T$_{\rm ex}>100$\,K, T$_{\rm
  ex}= 24.7\pm0.5$\,K and T$_{\rm ex}=52.5\pm6.3$\,K for {\it Z}, {\it
  Y} and {\it X} respectively (where the temperature calculated for
the Z component may be discrepant and likely caused by a problem with the
model fit to the line; see Fig. 1).  This supports the suggestion from
the $^{12}$CO SLED that both {\it X}, and especially {\it Z}, are hotter
than the more massive {\it Y}.

\subsection{\bf Physical Interpretation}
\label{sec:structure}

\citet{Swinbank10Nature} show that the rest-frame $\sim 250$-$\mu$m
emission from SMM\,J2135 is concentrated in four bright, star-forming
regions each of which has a radius of $\sim$100\,pc spread across a
more diffuse structure with a total extent of 2\,kpc in projection.
Following this we fitted the integrated dust SED of the source (see \S
\ref{sec:Mgass}) assuming that $\sim20$\% of the 870-$\mu$m flux is
coming from a relatively cool ($T_d\sim 30$\,K) extended dust
component with a predicted radius of $\sim1$\,kpc and $\sim80$\%
emerges from the four $\sim$100-pc radius, hotter ($T_d\sim 60$\,K)
and optically thick, clumps.  Our modelling of both the $^{12}$CO SLED
and the kinematically decomposed [C{\sc i}] line ratios supports the
presence of two phases in the ISM within this galaxy with a cool,
15--25\,K, phase and a hotter, 45--60\,K phase (although it is likely
that these actually represent a continuum of properties).  We also
find an enhancement in the [C{\sc i}]/$^{12}$CO ratio (likely due to
low metallicity and/or higher cosmic ray flux) which is associated
with the cool phase in this system. 
Our kinematic model identifies a component {\it Y} which is relatively
cool and contains the about half of the gas in the system, while the
hotter material is more closely tied to component {\it Z} (and to a
lesser extent {\it X}).  It is tempting to associate these kinematic
components with the clumps seen in the SMA map, but as we discuss
below, it is more likely that the clumps lie within these kinematic
structures, rather than having a unique one-to-one relationship.

Our observations also provide other limits on the characteristic size
and morphology for the emission regions. We can use the characteristic
density derived from the PDR analysis, together with the total gas
mass to derive an effective radius of the system.  Adopting
M(H$_2$+He)\,$\sim 4\times10^{10}$\,M$_\odot$ and a density of
$\sim10^{4}$\,cm$^{-3}$ from the \citet{Kaufman99} models, we derive
the radius of an equivalent sphere of $\sim300$\,pc (or $\sim 700$\,pc
for a disk with a thickness of $\sim 100$\,pc).  These sizes are larger
than the clump sizes in the sub-millimetre, which again is consistent with
the interpretation from the LVG modelling that the $^{12}$CO is much
more wide spread than the star-forming clumps.  If we similarly split
the gas mass into four equal clumps, then the expected sizes of these
are $\sim200$\,pc in radius, which is slightly larger than the far-infrared
sizes of the star-forming regions within SMM\,J2135.

We can also use the value of G$_{0}\sim10^{3.6}$ Habing fields, as
determined from the PDR models, to estimate the characteristic size of
the galaxy.  Assuming a Salpeter IMF and a constant SFR in the burst
then the ionising flux (6--13.6eV) is $\sim 50$\% of the bolometric
luminosity for ages of 10--100\,Myrs.  We can determine the size of a
region where the typical ionising field, as estimated from our
bolometric luminosity, equals G$_{0}$.  This gives an estimate for the
system's radius of $\sim 1$\,kpc, similar to the characteristic size
inferred from the LVG modelling of the dust SED and the overall extent
of the system from the SMA observations.

We can use the kinematics of the components and the potential
sizes and estimate the gas mass fractions in these various
components. We first note that the dynamical mass of a uniform sphere
with velocity dispersion of $\sigma \sim 75$--160\,km\,s$^{-1}$
(Table~3) and an effective radius of $r\sim200$--1000\,pc is
$\sim0.1$--$3\times10^{10}$\,M$_\odot$.  The gas masses from
Fig.~\ref{fig:CO_SLED_tot2} (using $\alpha=2.0$) are $\sim
7$--$23\times 10^9$\,M$_\odot$, similar to the range of dynamical
masses, indicating that the gas reservoirs in these components must
have effective radii comparable to the estimates above from the
various size indicators.  This also shows that we cannot
significantly increase $\alpha$, for example to the Milky Way value of
$\alpha \sim 4.6$, since this would result in a range of gas masses of
$\sim1.4$--$4.7\times10^{10}$\,M$_{\odot}$, in excess of the dynamical
masses unless the gas has disk-like kinematics and we are seeing these
almost face on, which is unlikely given the large velocity dispersion.

Our size estimates for the gas reservoirs are larger than the size of
the clumps in the SMA map and are more comparable to the extent of the
whole system, which suggests that the cold gas is more widely
distributed than the $\lsim 100$-pc radius far-infrared clumps.  Thus,
these clumps are likely to be knots of emission lying within a colder
and more extended structure. Although the clumps will trace the
kinematics of the gas in which they are embedded, they may not
represent the deepest parts of the potential well, and so the star
formation may be free to migrate around the gas reservoir.  These
intense star-forming regions have crossing times of just a few Myrs
and they have free-fall SFRs \citep{Krumholtz06} of $\sim 0.001$
indicating that they are supported against collapse (most likely by
internal turbulence).

Finally, with an estimate of the overall extent of the gas reservoir,
we investigate where SMM\,J2135 lies on the Kennicutt-Schmidt
relation, which links the star-formation- and gas surface density
($\Sigma_{\rm SFR}$ and $\Sigma_{\rm gas}$ respectively).  This
relation is approximately linear up to a threshold of
$\sim0.1$\,M$_{\odot}$\,pc$^{-2}$, above which the density of the ISM
becomes dominated by molecular gas and the star-formation law instead
follows $\Sigma_{\rm SFR}\propto\Sigma_{\rm gas}^{1.4}$ \citep{KS98}.
Previous work has shown broad universality of this relation with
redshift, although at $z\sim2$, both surface brightness and instrument
limitations mean that a direct comparison to low-redshift galaxies has
been difficult, with extrapolations from high-J CO in intrinsically
luminous galaxies providing most constraints
(e.g. \citealt{Genzel10}).  Some of this tension can be alleviated by
using the CO(1--0) emission (or LVG modelling of the full CO SLED) to
provide a more reliable estimate of the gas mass and spatial extent,
and in Fig.~\ref{fig:KS} we show the Kennicutt-Schmidt relation and
overlay the position of SMM\,J2135.  We also include on the plot the
high-redshift galaxies where CO(1--0) observations are available.
This figure shows that SMM\,J2135 lies at the high star-formation- and
gas-density end of the correlation, although within the scatter seen
for local ULIRGs.

Although we do not have spatially resolved spectroscopy to probe the
gas properties on $\sim100$\,pc scales, we can ask where the
individual star-forming regions would lie if they too followed the
Kennicutt-Schmidt relation.  Assuming that 80\% of the bolometric
luminosity arises in these clumps, and that each clump has a radius of
$r_{\rm SF}$=$r_{\rm gas}\sim100$\,pc, then the inferred gas density
would be $n\sim3\times10^5$\,cm$^{-2}$, which is substantially higher
than that inferred from the PDR and LVG modelling of the dense gas
tracers ($n\sim10^{3.5-4.5}$\,cm$^{-2}$).  However, taking the
remaining 20\% of the bolometric luminosity with a radius of 1\,kpc
(and assuming the gas lies in an extended disk), the density is
$n\sim0.2$--$1\times10^4$\,cm$^{2}$ (with the main uncertainty being
the disk thickness).  Thus, although this calculation should be
considered crude, the results are consistent with a model in which the
sub-millimetre emission predominantly traces the star-forming clumps
on $\sim100$\,pc scales, but these clumps are simply embedded in a
much more extended structure which dominates the $^{12}$CO emission.
These results may also imply a break-down of the Kennicutt-Schmidt
relation on scales of individual GMCs ($<100$\,pc, see also
\citealt{Onodera10}) although clearly high resolution millimetre
observations would accurately constrain their sizes, which is a
fundamental step in understanding the conditions within the ISM of
high-redshift galaxies (P. Cox et al. in prep).

\section{Conclusions }
\label{conc}

We have presented detections of a number of rest-frame, far-infrared
and sub-millimetre molecular and atomic emission lines in the spectrum
of the lensed sub-millimetre galaxy SMM\,J2135. We have used these
lines to construct the $^{12}$CO SLED up to $J_{\rm upper}=10$ and
analysed this and the other species, including the HCN(3--2) and [C{\sc
  i}] fine structure lines, to investigate the physical conditions
within the interstellar medium in this galaxy.  Our main results
are:\\

The $^{12}$CO SLED resembles that of local starburst galaxies.  An LVG
analysis of the $^{12}$CO SLED shows at least two phases with
temperatures $T\sim 25$ and 60\,K. 

The dust SED is best fit by a model in which the galaxy comprises an
extended component with radius $\sim1$--2\,kpc with $T_{d}\sim 30$\,K
and four high-density, star-forming regions with sizes of order $r\sim
200$\,pc and T$_{d}\sim 60$\,K.  These characteristics are roughly
consistent with the rest-frame $\sim 250$-$\mu$m morphology of the
system as seen by SMA and the two components may link to the phases
identified from the $^{12}$CO SLED.  At $J_{\rm upper}=1$--0, the
cooler, diffuse phase appears to comprise 50\% of the molecular gas.
We also derive a carbon mass of M$_{\rm CI} =(9.1 \pm 0.2) \times
10^{6}$\,M$_{\odot}$ and hence a carbon abundance of $\sim
4\times10^{-5}$ which is slightly above the Galactic value.  LVG
modelling suggests that the dense, star-forming regions have
abundances similar to that seen in the Milky Way, whilst the extended
diffuse gas has lower metallicity (or a higher cosmic ray flux).

Using the $^{12}$CO emission and assuming $\alpha=0.8$ we estimate the
total molecular gas mass from the $^{12}$CO(1--0) emission as
M(H$_2$)\,$=(1.4\pm0.1)\times 10^{10}$\,M$_{\odot}$. Comparing this
with that derived from $J_{\rm upper}\geq3$ transitions, assuming
thermalised gas, shows that these predicted gas masses are a factor
1.5--$3.6\times$ too low, indicative of multiple gas phases, as seen in
local starbursts.  However, our LVG analysis indicates a higher gas
mass, M(H$_2$)\,$\sim 4\times 10^{10}$\,M$_{\odot}$, yielding an
effective CO-to-H$_2$ conversion factor of $\alpha\sim 2.0$. If these
conditions are relevant in the wider SMG population then this suggests
that previous, $J_{\rm upper}\geq3$ $^{12}$CO observations of SMGs may
have underestimated the molecular gas content of SMGs by at least a
factor of $\sim 2$--$5\times$ (see also
\citealt{Ivison10L1L2,Harris10}).

Within SMM\,J2135 we detect the HCN(3--2) emission at the 3-$\sigma$
level, and use this to estimate a dense gas mass of M$_{\rm
  dense}$(H$_2$)\,$=\alpha_{\rm HCN}$\,L$'_{\rm HCN}<4.5 \times
10^{10}$\,M$_{\odot}$ assuming $\alpha_{\rm HCN}=10$ \citep{Gao04b}.
Within the uncertainties in $\alpha_{\rm HCN}$ this is consistent with
the gas mass derived from $^{12}$CO, indicating that HCN may be a
promising route to constraining the dense gas masses of high-redshift
galaxies in the future \citep{Gao07}, particularly if two or more HCN
lines are measured. We also find that the HCN/FIR ratio is consistent
with that of local ULIRGs and star-forming galaxies with no evidence
for AGN contribution to the FIR luminosity.

We use the atomic fine-structure and molecular line ratios to
investigate the excitation conditions within the ISM, and find that
they are similar to those found in local starbursts and ULIRGs.  Using
a grid of PDR models, we show that the molecular gas in SMM\,J2135 is
exposed to a UV radiation field which is $\sim1000\times$ more intense
than that of the Milky Way, and has a density of
$n\sim10^{4}$\,cm$^{-3}$, both characteristic of the central regions
of a typical starburst galaxy.  Thus, the inferred density and far-UV
radiation field strength of the ISM appear similar to those seen in
local ULIRGs consistent with the interpretation of SMGs as the
high-redshift analogs of these merger-driven starbursts.  However, the
spatial extent and mass of the gas reservoir (and spatial extent of
the star formation) is larger than typically found in local ULIRGs.

We find that the $^{12}$CO emission shows multiple kinematic components and we
decompose the SLEDs into three kinematic components to
investigate the excitation variations within the system.  We find
tentative evidence of temperature variation between the kinematic
components, although we defer a detailed discussion of the
kinematics to a future paper (P. Cox et al. in prep).

Using a number of approaches we have derived estimates for the
effective size of the gas reservoir within this system.  All of our
estimates are larger than the size of the clumps detected in the
870-$\mu$m map from SMA, implying that the gas traces a more extended
structure.  Overall, our modelling of the $^{12}$CO SLED supports the
presence of two phases in the ISM, with one phase associated with
$\sim$solar metallicity star-forming clumps with $>100$\,pc sizes,
embedded in a low metallicity (or high cosmic ray flux), extended
component with a characteristic size of $\sim1$\,kpc. However, even
the ``extended'' component has a much higher gas density than
typically observed in local starbursts such as M\,82 (although a
similar density to local ULIRGs such as Arp\,220), and is exposed to a
UV radiation field which is $\sim1000\times$ more intense than the
Milky Way \citep{Gerin98}.

Assuming that the molecular gas is evenly distributed, we show that
the integrated properties of the galaxy follow the Kennicutt-Schmidt
relation.  We also test where the individual star-forming regions
would lie if they too followed the Kennicutt-Schmidt relation, and we
show that on $\sim100$\,pc scales, the gas density would be
$\sim30\times$ higher than that inferred from the LVG or PDR analysis.
However, the properties of the diffuse gas disk suggest a
characteristic density $n\sim0.2$--$1\times10^4$\,cm$^{2}$, which is
comparable to that inferred from the LVG and PDR models.  These
results are consistent with a model in which the sub-millimetre
emission predominantly traces the dense cores of the star-forming
regions, but that these clumps are simply embedded in a much more
extended and diffuse structure which dominates the gas emission.
These results may also imply a break-down of the Kennicutt-Schmidt
relation on scales of individual GMCs (e.g.\ $<100$\,pc), although
clearly only high-resolution millimetre observations will accurately
constrain their sizes.

These results show that it is possible to unravel the complex ISM
physics within starburst galaxies at high redshift at a level of
detail which, until recently has only been possible in the $z<0.1$
Universe.  Our observations provide an unique window into the physics
of star-formation at $z\sim2$ with signal-to-noise that would
otherwise require the increased light grasp of the next generation
facilities. Indeed this source provides insight into the science that
will be routinely possible once ALMA reaches full science operations.

\section*{acknowledgments}

ALRD acknowledges an STFC studentship, AMS gratefully acknowledges a
Sir Norman Lockyer Royal Astronomical Society Fellowship, IRS
acknowledges support from STFC.  We are very grateful to Padeli
Papadopoulos for extensive disccusion and comments on this manuscript.
We also thank Estelle Bayet, Carsten Kramer, Rowin Meijerink, Serena
Viti, Paul van der Werf and Chris Wilson for a number of useful
discussions. We thank the anonymous referee for useful comments which
added to the clarity and content of this paper. The observations in
this paper were carried out with the IRAM 30\,m and the Plateau de
Bure Interferometer.  IRAM is supported by INSU/CNRS (France), MPG
(Germany) and IGN (Spain).  We also thank the ESO Director for
generously granting DDT observations with SHFI as part of program
283.A-5014 which is based on data acquired with the Atacama Pathfinder
Experiment (APEX).  APEX is a collaboration between the
Max-Planck-Institut fur Radioastronomie, the European Southern
Observatory, and the Onsala Space Observatory. The Submillimeter Array
is a joint project between the Smithsonian Astrophysical Observatory
and the Academia Sinica Institute of Astronomy and Astrophysics and is
funded by the Smithsonian Institution and the Academia Sinica.
Zpectrometer observations were taken as part of program 09A-040.
Zpectrometer is supported by National Science Foundation grants
AST-0503946 and AST-070865, and by the National Radio Astronomy
Observatory.  The National Radio Astronomy Observatory is a facility
of the National Science Foundation operated under cooperative
agreement by Associated Universities, Inc.

\bibliographystyle{mn2e}
\bibliography{ref}

\end{document}